\documentclass[amsmath,amsfonts,amssymb,prb,superscriptaddress,reprint,floatfix]{revtex4-1}
\usepackage{bm}
\usepackage{color}
\usepackage{graphicx}

\begin{document}
\title{On the Magnetic Structure of Density Matrices}
\author{Thomas M. Henderson}
\affiliation{Department of Chemistry, Rice University, Houston, Texas 77005, USA}
\affiliation{Department of Physics and Astronomy, Rice University, Houston, Texas 77005, USA}

\author{Carlos A. Jim\'enez-Hoyos}
\affiliation{Department of Chemistry, Wesleyan University, Middletown, CT 06459-0180, USA}

\author{Gustavo E. Scuseria}
\affiliation{Department of Chemistry, Rice University, Houston, Texas 77005, USA}
\affiliation{Department of Physics and Astronomy, Rice University, Houston, Texas 77005, USA}

\begin{abstract}
The spin structure of wave functions is reflected in the magnetic structure of the one-particle density matrix.  Indeed, for single determinants we can use either one to determine the other.  In this work we discuss how one can simply examine the one-particle density matrix to faithfully determine whether the spin magnetization density vector field is collinear, coplanar, or noncoplanar.  For single determinants, this test suffices to distinguish collinear determinants which are eigenfunctions of $\hat{S}_{\hat{n}}$ from noncollinear determinants which are not.  We also point out the close relationship between noncoplanar magnetism on the one hand and complex conjugation symmetry breaking on the other.  Finally, we use these ideas to classify the various ways single determinant wave functions break and respect symmetries of the Hamiltonian in terms of their one-particle density matrix.
\end{abstract}

\maketitle

\section{Introduction}
Magnetic structures are ubiquitous in nature and are of significant technological importance.  At the microscopic level, we associate magnetism with electronic or nuclear spin: the spin structure of the electronic wave function yields information about observed magnetic properties.  At the mean-field level, electronic magnetism is frequently associated with spin symmetry breaking, simply because most spin eigenfunctions cannot be described by a mean-field wave function.  We should note, however, that restricted open-shell wave functions can be spin eigenfunctions and yet have magnetic character.

On the one hand, the symmetry breaking of Hartree-Fock is certainly artificial: for finite systems, the exact solution does not break symmetries.  On the other hand, this symmetry breaking is not entirely unphysical, either.  For example, consider the dissociation of the H$_2$ molecule.  For large bond lengths, Hartree-Fock breaks spin symmetry, localizing the $\uparrow$-spin electron on one atom and the $\downarrow$-spin electron on the other.  While the exact solution is entangled and does not have broken spin symmetry, it is also true that the exact solution, unlike the symmetry-adapted Hartree-Fock, always has one electron on one atom and the other electron on the other atom (at infinite separation).  Thus, the broken spin symmetry has a certain degree of physical correctness: both the Hartree-Fock solution and the exact solutions display antiferromagnetism.  What the broken-symmetry mean-field lacks is entanglement; it gives a sort of classical picture of the dissociated limit.

The story, somewhat unfortunately, is slightly more complicated than that.  In addition to breaking $\hat{S}^2$ spin symmetry, Hartree-Fock can also break $\hat{S}_z$ spin symmetry, in what is known as generalized Hartree-Fock (GHF)\cite{fukutome1981,stuber2003,jimenezhoyos2011}.  But not all GHF solutions are alike.  Some may actually have an axis of spin quantization -- for example, the wave function may be an eigenstate of $\hat{S}_x$.  Though this would appear to be a GHF-type wave function, it is actually just an unrestricted Hartree-Fock (UHF) determinant with a rotated axis of spin quantization.  We can always create such a solution by acting a spin rotation operator on a UHF determinant.  But while these kinds of ``GHF'' solutions have a collinear (i.e. ferromagnetic or antiferromagnetic) structure, other kinds of GHF solutions may have coplanar but noncollinear magnetic structure, or even a general noncoplanar ordering, and when we refer to a GHF determinant we are really interested in one which is fully noncollinear.  These noncollinear GHF states are particularly prevalent in systems which exhibit spin frustration\cite{Yamaguchi1999}.

How are we to distinguish these various kinds of magnetic orderings of broken spin symmetry wave functions?  Conceptually this seems easy enough: one could simply plot the magnetization vector field defined in Eqn. \ref{Def:MagnetizationVector} below and examine it.  But this is by no means a practical solution except, perhaps, for lattice Hamiltonians.

An important step was provided by Small, Sundstrom, and Head-Gordon (SSHG)\cite{Small2015}.  They pointed out that because a collinear wave function is necessarily an eigenfunction of some spin operator $\hat{S}_{\hat{n}}$, where $\hat{n}$ is some spatial direction, one can determine whether a wave function is collinear or not by looking for the direction $\hat{n}$ which minimizes the fluctuation $\langle \hat{S}_{\hat{n}}^2 \rangle - \langle \hat{S}_{\hat{n}} \rangle^2$.  If in some direction the fluctuation vanishes, the wave function must be an $\hat{S}_{\hat{n}}$ eigenfunction.  This leads to the SSHG test to determine collinearity: a wave function is collinear if and only if the lowest eigenvalue of a matrix $\mathbf{A}$ vanishes, where the elements of $\mathbf{A}$ are
\begin{equation}
A_{ij} = \langle \hat{S}_i \, \hat{S}_j \rangle - \langle \hat{S}_i \rangle \, \langle \hat{S}_j \rangle
\label{Def:SSHGMatrix}
\end{equation}
and where $i$ and $j$ run over $x$, $y$, and $z$.  Note that this test in general requires the two-particle density matrix.  Moreover, it tests the spin structure of the wave function where we are more interested in examining the magnetic structure of the electronic density.  Obviously the former determines the latter, but testing the latter is, as we shall see, perhaps somewhat simpler in that it does not require the two-particle density matrix.  Finally, the SSHG test does not distinguish between coplanar and noncoplanar magnetizations, which would appear to arise from wave functions which break $\hat{S}_{\hat{n}}$ symmetry in different ways.

In this work, we seek to do several things.  First, we provide a test for the collinearity or noncollinearity of the magnetization density, based on the structure of the spinorbital one-particle density matrix.  This test is equivalent to the SSHG test for single determinant wave functions, though we provide an alternative conceptual motivation.  We also show how to distinguish between coplanar and noncoplanar magnetization densities; this test is motivated by the observation that a noncoplanar magnetization density requires a complex wave function, and is novel.  Finally, we note that while testing the magnetic structure of the one-particle density matrix does not allow us to infer too much about the spin characteristics of a general correlated wave function, they do allow us to determine whether a single determinant is collinear (i.e. an eigenfunction of $\hat{S}_{\hat{n}}$ for some $\hat{n}$) or not.  Since our test and that of SSHG are equivalent for single determinants, this is none too surprising, but it allows us to extend the work of Fukutome\cite{fukutome1981} and of Stuber and Paldus\cite{stuber2003}, who classified Hartree-Fock solutions in terms of the occupied molecular orbital coefficients. We show the equivalent classifications in terms of the one-particle density matrix.  We demonstrate our ideas for a handful of systems for which GHF solutions can be found.

\section{The Spinorbital One-Particle Density Matrix and the Magnetization Density}
Before we can discuss collinearity tests, we will require some preparatory material.

Let us begin, then, by considering the full spinorbital one-particle density matrix associated with a normalized state $|\Psi\rangle$, which may or may not be a single determinant and which we write as
\begin{equation}
\gamma^{\eta\xi}_{\mu \nu} = \langle \Psi| c_{\nu_\xi}^\dagger \, c_{\mu_\eta} |\Psi\rangle
\end{equation}
where $\nu$ and $\mu$ index spatial basis functions and $\eta$ and $\xi$ are spin indices.  Quite generally, in a spin-orbital basis in which the first block index corresponds to $\uparrow$ spin and the second block index corresponds to $\downarrow$ spin, we have
\begin{equation}
\bm{\gamma} = \begin{pmatrix} \bm{\gamma}^{\uparrow \uparrow} & \bm{\gamma}^{\uparrow \downarrow} \\ \bm{\gamma}^{\downarrow \uparrow} & \bm{\gamma}^{\downarrow \downarrow} \end{pmatrix}.
\end{equation}

For our purposes, it is more convenient to decompose the density matrix into a charge component $\mathbf{P}$ and spin components $\vec{\mathbf{M}}$ as
\begin{equation}
\bm{\gamma} = \begin{pmatrix} \mathbf{P} + \mathbf{M}^z & \mathbf{M}^x - \mathrm{i} \, \mathbf{M}^y \\ \mathbf{M}^x + \mathrm{i} \, \mathbf{M}^y & \mathbf{P} - \mathbf{M}^z \end{pmatrix} = \mathbf{P} \otimes \bm{1} + \vec{\mathbf{M}} \otimes \vec{\bm{\sigma}}.
\end{equation}
Here, $\bm{1}$ is the identity matrix in spinor space, $\vec{\bm{\sigma}}$ is the vector of Pauli matrices, and $\otimes$ denotes the Kronecker product; $\mathbf{P}$ is the charge density matrix and $\vec{\mathbf{M}}$ is the vector of spin density matrices:
\begin{equation}
\vec{\mathbf{M}} = \left( \mathbf{M}^x, \mathbf{M}^y, \mathbf{M}^z \right).
\end{equation}
The individual component matrices are
\begin{subequations}
\begin{align}
\mathbf{P} &= \frac{1}{2} \, \left(\bm{\gamma}^{\uparrow\uparrow} + \bm{\gamma}^{\downarrow\downarrow}\right),
\\
\mathbf{M}^x &= \frac{1}{2} \, \left(\bm{\gamma}^{\downarrow\uparrow} + \bm{\gamma}^{\uparrow\downarrow}\right),
\\
\mathbf{M}^y &= \frac{1}{2 \, \mathrm{i}} \, \left(\bm{\gamma}^{\downarrow\uparrow} - \bm{\gamma}^{\uparrow\downarrow}\right),
\\
\mathbf{M}^z &= \frac{1}{2} \, \left(\bm{\gamma}^{\uparrow\uparrow} - \bm{\gamma}^{\downarrow\downarrow}\right),
\end{align}
\end{subequations}
and can be extracted from
\begin{subequations}
\begin{align}
P_{\mu\nu} &= \frac{1}{2} \, \langle \Psi| c_{\nu_\uparrow}^\dagger \, c_{\mu_\uparrow} + c_{\nu_\downarrow}^\dagger \ c_{\mu_\downarrow} |\Psi\rangle \equiv \langle \Psi| \hat{P}_{\mu\nu} |\Psi\rangle,
\\
M^x_{\mu\nu} &= \frac{1}{2} \, \langle \Psi| c_{\nu_\uparrow}^\dagger \, c_{\mu_\downarrow} + c_{\nu_\downarrow}^\dagger \ c_{\mu_\uparrow} |\Psi\rangle \equiv \langle \Psi| \hat{M}^x_{\mu\nu} |\Psi\rangle,
\\
M^y_{\mu\nu} &= \frac{1}{2 \, \mathrm{i}} \, \langle \Psi| c_{\nu_\uparrow}^\dagger \, c_{\mu_\downarrow} - c_{\nu_\downarrow}^\dagger \ c_{\mu_\uparrow} |\Psi\rangle \equiv \langle \Psi| \hat{M}^y_{\mu\nu} |\Psi\rangle,
\\
M^z_{\mu\nu} &= \frac{1}{2} \, \langle \Psi| c_{\nu_\uparrow}^\dagger \, c_{\mu_\uparrow} - c_{\nu_\downarrow}^\dagger \ c_{\mu_\downarrow} |\Psi\rangle \equiv \langle \Psi| \hat{M}^z_{\mu\nu} |\Psi\rangle.
\end{align}
\end{subequations}

Having defined these magnetization density matrices, we can now define the magnetization vector field or, if one prefers, the spin density vector field.  Choosing our basis to be real as we can do without loss of generality, the magnetization vector at a point in space is simply
\begin{equation}
\vec{m}(\vec{r}) = \sum_{\mu,\nu} \chi_\mu(\vec{r}) \, \chi_\nu(\vec{r}) \, \vec{\mathbf{M}}_{\mu\nu}.
\label{Def:MagnetizationVector}
\end{equation}
Note that only the symmetric part of $\vec{\mathbf{M}}$ contributes to the magnetization vector. Because $\vec{\mathbf{M}}$ is Hermitian, its symmetric part is its real part.  If the density matrix $\bm{\gamma}$ is real, then $\mathbf{M}^y$ is purely imaginary, hence $m_y(\vec{r})$ vanishes identically and the magnetization density $\vec{m}(\vec{r})$ is coplanar.  In other words, a real wave function has coplanar magnetism.  The converse is not necessarily true: coplanar magnetism does not necessarily imply a real wave function.  Conceivably, we could have, for example, $\mathbf{M}^y$ purely imaginary with complex $\mathbf{M}^x$ and $\mathbf{M}^z$.  We should note that the imaginary parts of of $\vec{\mathbf{M}}$ do contribute to the spin current density\cite{fukutome1981}.  We shall have more to say on the spin current density later.

\section{Testing Magnetic Structure}
Suppose that a wave function is an eigenfunction of $\hat{S}_z$ (and of the total number operator).  Then that wave function has a definite number of $\uparrow$-spin and of $\downarrow$-spin electrons.  For such a wave function, $\gamma^{\uparrow\downarrow}$ and $\gamma^{\downarrow\uparrow}$ must be identically zero, because the operator $c^\dagger_\uparrow \, c_\downarrow$ changes the number of electrons of each spin direction when acting on that wave function.  Thus, an $\hat{S}_z$ eigenfunction has a block diagonal spinorbital density matrix, and if the spinorbital density matrix is not block diagonal, the wave function is not an eigenfunction of $\hat{S}_z$.

Note that if the spinorbital density matrix is block diagonal, we cannot guarantee that the underlying wave function is an eigenfunction of $\hat{S}_z$ unless the wave function is a single determinant.  If the wave function is a single determinant, diagonalizing the spinorbital density matrix allows us to obtain the occupied orbitals; if the density matrix is block diagonal, the occupied orbitals can be chosen to be $\hat{S}_z$ eigenfunctions, and if the occupied orbitals are $\hat{S}_z$ eigenfunctions, so is the determinant.  To test the spin structure of a general wave function, we require the two-particle density matrix, as SSHG pointed out.

We have discussed the special case of magnetization aligned along the $z$ axis, but of course nothing privileges that axis.  Quite generally, if a wave function is an $\hat{S}_{\hat{n}}$ eigenfunction then the spinorbital density matrix is block diagonal in spin blocks where $\uparrow$ and $\downarrow$ are defined relative to $\hat{n}$.  If the density matrix cannot be brought to this form, the magnetization is noncollinear and the wave function is not an eigenfunction of $\hat{S}_{\hat{n}}$ for any direction $\hat{n}$; if the density matrix can be brought to this form, the magnetization vector field is collinear and the wave function, if a single determinant, is guaranteed to be an eigenfunction of $\hat{S}_{\hat{n}}$.

\subsection{Spin Rotation Operators}
To test this possibility, we must consider spin rotation operators.  We define a unitary spin rotation operator\cite{PHF}
\begin{equation}
R(\Omega) = \mathrm{e}^{\mathrm{i} \, \gamma \, \hat{S}_z} \, \mathrm{e}^{\mathrm{i} \, \beta \, \hat{S}_y} \, \mathrm{e}^{\mathrm{i} \, \alpha \, \hat{S}_z}.
\label{Def:RotationOperators}
\end{equation}
where $\Omega$ stands for the collection of rotation angles $\left(\alpha,\beta,\gamma\right).$   The three angles $\alpha$, $\beta$, and $\gamma$ are Euler angles and cover the sphere.  With this operator we can define a rotated state
\begin{equation}
|\tilde{\Psi}_\Omega\rangle = \hat{R}(\Omega) |\Psi\rangle.
\end{equation}
Note that if the Hamiltonian commutes with the spin operators then $|\Psi\rangle$ and $|\tilde{\Psi}_\Omega\rangle$ are degenerate
\begin{equation}
\langle \tilde{\Psi}_\Omega | \hat{H} | \tilde{\Psi}_\Omega \rangle
          = \langle \Psi | \hat{R}^\dagger(\Omega) \, \hat{H} \, \hat{R}(\Omega) |\Psi\rangle
          = \langle \Psi| \hat{H} |\Psi\rangle
\end{equation}
where we have used the fact that $\hat{H}$ commutes with $\hat{R}$ and that $\hat{R}^\dagger \, \hat{R} = 1$.  Note also that if $|\Psi\rangle$ is a single determinant, so too is $|\tilde{\Psi}(\Omega)\rangle$, because $\hat{R}$ is a series of exponentials of one-body operators, i.e. it is a Thouless transformation\cite{Thouless1960}.  Together, these observations imply that if $|\Psi\rangle$ is a solution of the Hartree-Fock equations, then so too is $|\tilde{\Psi}(\Omega)\rangle$.  In fact, the collection of states $|\tilde{\Psi}(\Omega)\rangle$ forms a manifold known as the Goldstone manifold and is used in spin symmetry projection\cite{Ring80,Blaizot85,PHF}.

Let us now consider the rotated density matrix.  Generically, we will have
\begin{subequations}
\begin{align}
\tilde{\gamma}^{\eta\xi}_{\mu \nu} 
 &= \langle \tilde{\Psi}| c_{\nu_\xi}^\dagger \, c_{\mu_\eta} |\tilde{\Psi}\rangle
\\
 &= \langle \Psi| \hat{R}^\dagger \, c_{\nu_\xi}^\dagger \, c_{\mu_\eta} \, \hat{R}|\Psi\rangle
\\
 &= \langle \Psi| \hat{R}^\dagger \, c_{\nu_\xi}^\dagger \, \hat{R} \, \hat{R}^\dagger \, c_{\mu_\eta} \, \hat{R}|\Psi\rangle
\\
 &= \langle \Psi| \tilde{c}_{\nu_\xi}^\dagger \, \tilde{c}_{\mu_\eta} |\Psi\rangle
\end{align}
\end{subequations}
where $\tilde{c}$ is the rotated annihilation operator.  The first line shows a sort of active rotation perspective: the rotation operator is understood as rotating the wave function, and we consider the density matrix expressed in terms of the original spin coordinates.  We see, however, from the last line that this is equivalent to a passive rotation perspective: the wave function is left alone and the underlying basis is rotated.  This latter perspective is more helpful for our purposes: if by such a rotation we can eliminate $\mathbf{M}^x$ and $\mathbf{M}^y$, the magnetization density is collinear.

Using the representation of $\hat{S}_z$ in terms of fermionic creation and annihilation operators,
\begin{equation}
\hat{S}_z = \frac{1}{2} \, \left(c_\uparrow^\dagger \, c_\uparrow - c_\downarrow^\dagger \, c_\downarrow\right),
\end{equation}
we see that
\begin{subequations}
\begin{align}
[c_\uparrow^\dagger, \hat{S}_z] &= -\frac{1}{2} \, c_\uparrow^\dagger,
\\
[c_\downarrow^\dagger, \hat{S}_z] &= \frac{1}{2} \, c_\downarrow^\dagger.
\end{align}
\end{subequations}
Then we can resum the commutator expansion analytically, and one can show that
\begin{subequations}
\begin{align}
\mathrm{e}^{-\mathrm{i} \, \theta \, \hat{S}_z} \, c_\uparrow^\dagger \, \mathrm{e}^{\mathrm{i} \, \theta \, \hat{S}_z}
 &= \mathrm{e}^{-\frac{1}{2} \, \mathrm{i} \, \theta} \, c_\uparrow^\dagger,
\\
\mathrm{e}^{-\mathrm{i} \, \theta \, \hat{S}_z} \, c_\downarrow^\dagger \, \mathrm{e}^{\mathrm{i} \, \theta \, \hat{S}_z}
 &= \mathrm{e}^{\frac{1}{2} \, \mathrm{i} \, \theta} \, c_\downarrow^\dagger.
\end{align}
\end{subequations}
This is turn implies that
\begin{subequations}
\begin{align}
\tilde{\mathbf{P}} &= \mathbf{P},
\\
\tilde{\mathbf{M}}^x &= \cos(\theta) \, \mathbf{M}^x + \sin(\theta) \, \mathbf{M}^y,
\\
\tilde{\mathbf{M}}^y &= \cos(\theta) \, \mathbf{M}^y - \sin(\theta) \, \mathbf{M}^x,
\\
\tilde{\mathbf{M}}^z &= \mathbf{M}^z.
\end{align}
\end{subequations}
where $\tilde{\mathbf{P}}$ and $\tilde{\mathbf{M}}^i$ are the components of the rotated density matrix $\tilde{\bm{\gamma}}$.  We can express this concisely as
\begin{subequations}
\begin{align}
\tilde{\mathbf{P}} &= \mathbf{P},
\\
\tilde{\mathbf{M}} &= \mathbf{R}_z(\theta) \, \mathbf{M}.
\end{align}
\end{subequations}
Here, $\mathbf{R}_z(\theta)$ is the rotation matrix corresponding to rotation by angle $\theta$ about the $z$ axis and $\tilde{\mathbf{M}}$ and $\mathbf{M}$ are written as column vectors.  One finds equivalent results for $\hat{S}_x$ and $\hat{S}_y$ spin rotations.  Spin rotations of the wave function manifest as spatial rotations of the magnetization density matrices.

Note that we are using the word ``spatial'' here in a somewhat cavalier sense: the directions $x$, $y$, and $z$ in the magnetization density matrices are not physically significant in the absence of an external field.

It may prove useful to note the spatial rotation matrices corresponding to the spin rotation operator of Eqn. \ref{Def:RotationOperators}.  We have
\begin{subequations}
\begin{align}
\mathbf{R}_x(\theta) &= \begin{pmatrix} 1 & 0 & 0 \\ 0 & \cos(\theta) & \sin(\theta) \\ 0 & -\sin(\theta) & \cos(\theta) \end{pmatrix},
\\
\mathbf{R}_y(\theta) &= \begin{pmatrix} \cos(\theta) & 0 & -\sin(\theta) \\ 0 & 1 & 0 \\ \sin(\theta) & 0 & \cos(\theta) \end{pmatrix},
\\
\mathbf{R}_z(\theta) &= \begin{pmatrix} \cos(\theta) & \sin(\theta) & 0 \\ -\sin(\theta) & \cos(\theta) & 0 \\ 0 & 0 & 1\end{pmatrix}.
\end{align}
\end{subequations}
Note that this is opposite the usual convention for passive rotations.  We have included $\mathbf{R}_x(\theta)$ for completeness.

Let us make one final observation.  While spin rotations cannot convert a collinear density matrix into a noncollinear density matrix or a coplanar density matrix into a noncoplanar one, they can convert a real density matrix into a complex density matrix.  The spin rotation operator, that is, does not commute with the complex conjugation operator defined below.  A complex conjugation eigenfunction, upon spin rotation, may cease to be a complex conjugation eigenfunction.

\subsection{Testing Collinearity and Coplanarity}
We can take advantage of the correspondence between spin rotations of $|\Psi\rangle$ and spatial rotations of $\vec{\mathbf{M}}$ to test the magnetic structure of the density matrix.  We note the following:
\begin{itemize}
\item If the spin density matrices $\vec{\mathbf{M}}$ are all identically zero, then the magnetization density vanishes.  If the wave function is a single determinant, it is an eigenfunction of $\hat{S}^2$ with eigenvalue zero, and is also an eigenfunction of $\hat{S}_{\hat{n}}$ with eigenvalue 0 for all directions $\hat{n}$.  This is the case for RHF.
\item If the spin density matrices can be rotated so that $\mathbf{M}^z$ is nonzero but both $\mathbf{M}^x$ and $\mathbf{M}^y$ are zero, then the magnetization density is collinear.  The underlying wave function is not a singlet (but may be an eigenfunction of $\hat{S}^2$).  If the wave function is a single determinant, it is definitely an eigenfunction of $\hat{S}_{\hat{n}}$ for some direction $\hat{n}$.  This is the case of UHF and also of rotated UHF solutions.
\item Otherwise the magnetization is noncollinear and the wave function is not an eigenfunction of $\hat{S}_{\hat{n}}$ for any direction $\hat{n}$, whether the wave function is a single determinant or not.  If the wave function is a single determinant, it is not an eigenfunction of $\hat{S}^2$.  This is the case of GHF.
\end{itemize}
In other words, if the wave function yields a nonzero spin density matrix, it is not a singlet; if the density matrix can be rotated to have the UHF structure, then the magnetization vector field is collinear and the wave function, if a single determinant, is definitely an eigenfunction of $\hat{S}_{\hat{n}}$; if the density matrix cannot be rotated to have the UHF structure then the magnetization vector field is noncollinear and the wave function is not an eigenfunction of $\hat{S}_{\hat{n}}$.

To see whether a spin density matrix vanishes or not, it is simplest to test its Frobenius norm.  Recall that the (square of the) Frobenius norm of a matrix $\mathbf{X}$ is
\begin{equation}
\| \mathbf{X}\|^2 = \sum_{pq} X_{pq} \, X_{pq}^\star = \mathrm{Tr}(\mathbf{X} \, \mathbf{X}^\dagger).
\end{equation}
Our matrices are Hermitian, so 
\begin{equation}
\| \mathbf{X} \|^2 = \mathrm{Tr}(\mathbf{X}^2).
\end{equation}

We wish to maximize the norm of one component of $\vec{\mathbf{M}}$.  To do so, we can diagonalize a matrix $\mathbf{T}$ given by
\begin{equation}
T_{ij} = \mathrm{Tr}(\mathbf{M}^i \, \mathbf{M}^j).
\label{Eqn:DefTMatrix}
\end{equation}
This matrix is real and symmetric.  Its diagonal components are the norms of the various magnetization density matrices.  Its off-diagonal components can be brought to zero by a sequence of rotations or, more correctly, we can bring $\mathbf{T}$ to diagonal form using spin rotation operators of the sort given in Eqn. \ref{Def:RotationOperators}.  Diagonalizing $\mathbf{T}$ is tantamount to finding the spin rotation which maximizes the norm of the largest component of $\vec{\mathbf{M}}$ and minimizes the norm of the smallest compotent; in other words, the diagonal elements of $\mathbf{T}$ cannot be rotated to be larger than the largest eigenvalue of $\mathbf{T}$ or smaller than the smallest eigenvalue of $\mathbf{T}$.

Our procedure in full is thus simple.  We build the matrix $\mathbf{T}$ and diagonalize it.  If $\mathbf{T}$ has three zero eigenvalues, then $\vec{\mathbf{M}}$ vanishes and the wave function, if a single determinant, has the RHF structure.  If $\mathbf{T}$ has two zero eigenvalues, the magnetization was collinear.  Otherwise it was noncollinear.  Depending on the outcome of the test and on whether the wave function is a single determinant or not, we may or may not be able to say whether the wave function itself is an eigenfunction of $\hat{S}^2$ or of $\hat{S}_{\hat{n}}$.  For single determinants, the test is equivalent to the test of SSHG (see below); for multideterminantal wave functions it is not.

If the magnetization is noncollinear, we can repeat the test but with a modified matrix
\begin{equation}
\mathcal{T}_{ij} = \mathrm{Tr}[\mathrm{Re}{(\mathbf{M}^i)} \, \mathrm{Re}{(\mathbf{M}^j)}].
\end{equation}
If there are any zero eigenvalues, the magnetization was coplanar because we could rotate to make one of the components of $\vec{\mathbf{M}}$ purely imaginary.  Note that this coplanarity test is new.

In a non-orthonormal basis,  we have
\begin{equation}
T_{ij} = \mathrm{Tr}(\mathbf{M}_i \, \mathbf{S} \, \mathbf{M}_j \, \mathbf{S})
\end{equation}
where $\mathbf{S}$ is the overlap matrix of spatial orbitals, and analogously for $\mathcal{T}_{ij}$.

Let us make a few caveats.  First, it is possible in principle for $\mathbf{T}$ to have one zero eigenvalue, which means we can bring $\vec{\mathbf{M}}$ to the form $\vec{\mathbf{M}} = (\mathbf{M}^x,\mathbf{0},\mathbf{M}^z)$.  This would of course correspond to the coplanar case, but where a typical coplanar magnetic structure has coplanar spin density but may have noncoplanar spin density matrices, this case corresponds to coplanar spin density matrices.  Second, we must point out the existence of paired UHF and paired GHF solutions (see below).  In these cases, $\vec{m}$ vanishes, and $\bm{\mathcal{T}} = \mathbf{0}$, yet $\mathbf{T}$ may have one or more nonzero eigenvalues.  Lastly, after diagonalization of $\mathbf{T}$ we choose directions such that $\| \mathbf{M}^y \| \le \| \mathbf{M}^x \| \le \| \mathbf{M}^z\|$, which we can always do as a matter of convenience.

\subsubsection{Collinear Spin Densities}
Let us make a quick comment on the collinear case.  If after the final rotation the density matrices are 
\begin{equation}
\vec{\mathbf{M}}^{\prime\prime\prime} = (\mathbf{0},\mathbf{0},\mathbf{Z}),
\end{equation}
then before that final rotation (i.e. after the second rotation) the density matrices must also have been
\begin{equation}
\vec{\mathbf{M}}^{\prime\prime} = (\mathbf{0},\mathbf{0},\mathbf{Z}).
\end{equation}
This in turn means that before the $y$ rotation (and therefore after the first $z$ rotation) the density matrices were 
\begin{equation}
\vec{\mathbf{M}}^\prime = (\mathbf{Z} \, \sin(\beta), \mathbf{0}, \mathbf{Z} \, \cos(\beta)).
\end{equation}
And lastly, this in turn means the initial unrotated density matrices were
\begin{equation}
\vec{\mathbf{M}} = (\mathbf{Z} \, \sin(\beta) \, \cos(\alpha),-\mathbf{Z} \, \sin(\beta) \, \sin(\alpha),\mathbf{Z} \, \cos(\beta)).
\end{equation}
A collinear solution, in other words, is characterized by a density matrix vector which is a spatial unit vector times a single matrix:
\begin{equation}
\vec{\mathbf{M}} = \hat{n} \, \mathbf{Z}
\end{equation}
where
\begin{equation}
\hat{n} = (\sin(\beta) \, \cos(\alpha), -\sin(\beta) \, \sin(\alpha), \cos(\beta)).
\end{equation}
An alternative test for whether a set of spin density matrices is collinear, then, is simply to see whether the components $\mathbf{M}^x$, $\mathbf{M}^y$, and $\mathbf{M}^z$ are all multiples of the same matrix $\mathbf{M}$.

\subsubsection{The SSHG Test}
The collinearity test of SSHG in general looks at eigenvalues of the matrix $\mathbf{A}$ defined in Eqn. \ref{Def:SSHGMatrix}.  Let us take a moment to rewrite this matrix for the case of a single determinant, using the language of the previous section.  We will assume an orthonormal basis for simplicity.

In their paper, SSHG note that for a single determinant, one finds
\begin{equation}
A_{ij} = -\mathrm{Tr}(\mathbf{O}_i \, \mathbf{O}_j) + \frac{1}{4} \, \delta_{ij} \, N,
\end{equation}
where $N$ is the number of electrons and
\begin{equation}
\mathbf{O}_i = \frac{1}{2} \, \mathbf{C}_\mathrm{occ}^\dagger \, \left(\bm{1} \otimes \bm{\sigma}_i\right) \, \mathbf{C}_\mathrm{occ}.
\end{equation}
Here, $\mathbf{C}_\mathrm{occ}$ is the matrix of occupied orbital coefficients.

Using the cyclic property of traces, we can equivalently write
\begin{align}
A_{ij} = -\frac{1}{4} \, \mathrm{Tr}[&\mathbf{C}_\mathrm{occ} \, \mathbf{C}_\mathrm{occ}^\dagger \, \left(\bm{1} \otimes \bm{\sigma}_i\right) 
\\
 & \times \mathbf{C}_\mathrm{occ} \, \mathbf{C}_\mathrm{occ}^\dagger \, \left(\bm{1} \otimes \bm{\sigma}_j\right)] + \frac{1}{4} \, \delta_{ij} \, N.
\nonumber
\end{align}
One can recognize the density matrix $\bm{\gamma} = \mathbf{C}_\mathrm{occ} \, \mathbf{C}_\mathrm{occ}^\dagger$, and note that
\begin{equation}
N = \mathrm{Tr}(\bm{\gamma}) = 2 \, \mathrm{Tr}(\mathbf{P}).
\end{equation}
Then one has
\begin{equation}
A_{ij} = -\frac{1}{4} \, \mathrm{Tr}[\bm{\gamma} \, \left(\bm{1} \otimes \bm{\sigma}_i\right) \, \bm{\gamma} \, \left(\bm{1} \otimes \bm{\sigma}_j\right)] + \frac{1}{2} \, \delta_{ij} \, \mathrm{Tr}(\mathbf{P}).
\end{equation}

Inserting our decomposition of $\bm{\gamma}$, one finds that the components of $\mathbf{A}$ are
\begin{equation}
A_{ij} = -\mathrm{Tr}(\mathbf{M}^i \, \mathbf{M}^j) + \frac{1}{2} \, \delta_{ij} \, \mathrm{Tr}[\mathbf{P} - \mathbf{P}^2 + \vec{\mathbf{M}} \cdot \vec{\mathbf{M}}].
\end{equation}

For a single determinant, $\bm{\gamma}$ is idempotent.  We have
\begin{equation}
\bm{\gamma}^2 = \left(\mathbf{P} \otimes \bm{1} + \mathbf{M}^i \otimes \bm{\sigma}^i\right)^2 = \mathbf{P} \otimes \bm{1} + \mathbf{M}^i \otimes \bm{\sigma}^i
\end{equation}
where here we employ the summation convention.  Using
\begin{equation}
\bm{\sigma}^i \, \bm{\sigma}^j = \delta_{ij} \, \bm{1} + \mathrm{i} \, \epsilon_{ijk} \, \bm{\sigma}^k,
\end{equation}
we see that the portion of $\bm{\gamma}^2$ which is proportional to the identity in spin space is simply $\mathbf{P}^2 + \mathbf{M}^i \, \mathbf{M}^i$.  Idempotency of the one-particle density matrix implies that
\begin{equation}
\mathbf{P} - \mathbf{P}^2 = \vec{\mathbf{M}} \cdot \vec{\mathbf{M}}.
\end{equation}
We can thus write the matrix $\mathbf{A}$ simply as
\begin{equation}
A_{ij} =- T_{ij}  + \delta_{ij} \, \mathrm{Tr}(\mathbf{P} - \mathbf{P}^2).
\end{equation}
For a single determinant, the SSHG test can be reformulated in terms of diagonalization of the simple matrix $\mathbf{T}$.

For the sake of completeness, we reiterate the three possibilities for a single determinant here, in terms of $\mathbf{T}$ and of $\mathbf{A}$:
\begin{itemize}
\item If the determinant is a singlet, then $\mathbf{A} = \mathbf{T} = \mathbf{0}$; both matrices of course have three zero eigenvalues.
\item If the determinant is collinear, then $\mathbf{T}$ has one non-zero eigenvalue $\lambda = \mathrm{Tr}(\mathbf{P} - \mathbf{P}^2)$ and two zero eigenvalues; the eigenvalues of $\mathbf{A}$ are $(0,\lambda,\lambda)$.
\item If the determinant is noncollinear, then $\mathbf{T}$ has no more than one zero eigenvalue (and usually has none); $\mathbf{A}$ has no zero eigenvalues.
\end{itemize}

\begin{table*}[t]
\caption{Classification of Hartree-Fock solutions in terms of preserved symmetries.  We show the names suggested by Fukutome and those suggested by Stuber for each of these solutions, and also include the structure of the matrix of orbital coefficients and any constraints placed on it as well as the structures of the charge and spin density matrices and constraints placed upon them.
\label{Tab:Symmetries}}
\begin{ruledtabular}
\begin{tabular}{lllll}
Fukutome     &  Stuber-Paldus  &  Symmetries  &  Structure of Occupied Orbital   & Structure of    \\
Designation  &  Designation    &  Preserved   &  Coefficient Matrix $\mathbf{C}_\mathrm{occ}$ & Density Matrices\\
\hline
TICS\footnote{Time-reversal Invariant Closed-Shell}      &
Real RHF                                                 &
$\hat{S}^2$, $\hat{S}_z$, $\hat{K}$, $\hat{\Theta}$      &
$\begin{pmatrix}
  \mathbf{C}_{\sigma\sigma}  & \bm{0}   \\
  \bm{0}  & \mathbf{C}_{\sigma\sigma}
\end{pmatrix},
\mathbf{C}_\mathrm{occ} \in \mathbb{R}$      &
$\vec{\mathbf{M}} = \vec{\bm{0}}$, $\mathbf{P} \in \mathbb{R}$
\\
CCW\footnote{Charge Current Wave}     &
Complex RHF                           &
$\hat{S}^2$, $\hat{S}_z$              &
$\begin{pmatrix}
  \mathbf{C}_{\sigma\sigma}  & \bm{0}   \\
  \bm{0}  & \mathbf{C}_{\sigma\sigma}
\end{pmatrix}$    &
$\vec{\mathbf{M}} = \vec{\bm{0}}$
\\
ASCW\footnote{Axial Spin Current Wave}   &
Paired UHF                               &
$\hat{S}_z$, $\hat{\Theta}$              &
$\begin{pmatrix}
  \mathbf{C}_{\sigma\sigma}  & \bm{0}   \\
  \bm{0}  & \mathbf{C}_{\sigma\sigma}^\star
\end{pmatrix}$  &
$\mathbf{M}^x = \mathbf{M}^y = \bm{0}$, $(\mathbf{P},\mathrm{i} \, \mathbf{M}^z) \in \mathbb{R}$
\\
ASDW\footnote{Axial Spin Density Wave}  &
Real UHF                                &
$\hat{S}_z$, $\hat{K}$                  &
$\begin{pmatrix}
  \mathbf{C}_{\sigma\sigma}  & \bm{0}   \\
  \bm{0}  & \mathbf{C}_{\sigma'\sigma'}
\end{pmatrix},
\mathbf{C}_\mathrm{occ} \in \mathbb{R}$  &
$\mathbf{M}^x = \mathbf{M}^y = \bm{0}$, $(\mathbf{P},\mathbf{M}^z) \in \mathbb{R}$
\\
ASW\footnote{Axial Spin Wave}           &
Complex UHF                             &
$\hat{S}_z$                             &
$\begin{pmatrix}
  \mathbf{C}_{\sigma\sigma}  & \bm{0}   \\
  \bm{0}  & \mathbf{C}_{\sigma'\sigma'}
\end{pmatrix}$   &
$\mathbf{M}^x = \mathbf{M}^y = \bm{0}$
\\
TSCW\footnote{Torsional Spin Current Wave}   &
Paired GHF                                   &
$\hat{\Theta}$                               &
$\begin{pmatrix}
\mathbf{C}_{\sigma \sigma}   & \mathbf{C}_{\sigma \sigma'}  \\
-\mathbf{C}_{\sigma\sigma'}^\star  & \mathbf{C}_{\sigma\sigma}^\star
\end{pmatrix}$  &
$(\mathbf{P},\mathrm{i} \vec{\mathbf{M}}) \in \mathbb{R}$
\\
TSDW\footnote{Torsional Spin Density Wave}    &
Real GHF                                      &
$\hat{K}$                                     &
$\begin{pmatrix}
\mathbf{C}_{\sigma\sigma}  & \mathbf{C}_{\sigma\sigma'}  \\
\mathbf{C}_{\sigma'\sigma}  & \mathbf{C}_{\sigma'\sigma'}
\end{pmatrix},
\mathbf{C}_\mathrm{occ} \in \mathbb{R}$  &
$(\mathbf{P},\vec{\mathbf{M}}) \in \mathbb{R}$
\\
TSW\footnote{Torsional Spin Wave}           &
Complex GHF                                 &
                                            &
$\begin{pmatrix}
\mathbf{C}_{\sigma\sigma}  & \mathbf{C}_{\sigma\sigma'}  \\
\mathbf{C}_{\sigma'\sigma}  & \mathbf{C}_{\sigma'\sigma'}
\end{pmatrix}$
\\
\end{tabular}
\end{ruledtabular}
\end{table*}

\section{Classification of Hartree-Fock Solutions}
We have seen that the collinearity test informs us about the symmetries of single determinants.  Let us therefore take a moment to revisit the classification of Hartree-Fock solutions in terms of symmetries, first proposed by Fukutome\cite{fukutome1981}  and later analyzed by Stuber and Paldus\cite{stuber2003}.  The various classifications are presented in Tab. \ref{Tab:Symmetries}.  In addition to the spin operators $\hat{S}^2$ and $\hat{S}_z$ we also have the complex conjugation operator $\hat{K}$ and the time reversal operator $\hat{\Theta}$.  For our purposes it is enough to define $\hat{K}$ and $\hat{\Theta}$ by their action on a single determinant.  Suppose a determinant $\Phi$ is specified by a matrix of occupied molecular orbital coefficients
\begin{equation}
\mathbf{C}_\mathrm{occ}(\Phi) = \begin{pmatrix} \mathbf{C}_\mathrm{occ}^\uparrow \\ \mathbf{C}_\mathrm{occ}^\downarrow \end{pmatrix}.
\end{equation}
Then the determinants $\hat{K} \Phi$ and $\hat{\Theta} \Phi$ are specified by matrices of occupied molecular orbital coefficients which are respectively
\begin{subequations}
\begin{align}
\mathbf{C}_\mathrm{occ}(\hat{K} \Phi)
 &= \begin{pmatrix} \left(\mathbf{C}_\mathrm{occ}^\uparrow\right)^\star \\ \left(\mathbf{C}_\mathrm{occ}^\downarrow\right)^\star \end{pmatrix},
\\
\mathbf{C}_\mathrm{occ}(\hat{\Theta} \Phi)
 &= \begin{pmatrix} -\left(\mathbf{C}_\mathrm{occ}^\downarrow\right)^\star \\ \left(\mathbf{C}_\mathrm{occ}^\uparrow\right)^\star \end{pmatrix}.
\end{align}
\end{subequations}
Thus, we have
\begin{equation}
\hat{\Theta} = -\mathrm{i} \, \bm{\sigma}^y \, \hat{K}.
\end{equation}
The classifications in Tab. \ref{Tab:Symmetries} were presented originally in terms of occupied molecular orbital coefficients; here, we list the corresponding constraints on the density matrix components, which were also discussed earlier by Weiner and Trickey\cite{Weiner1998}.

For the most part the constraints on the density matrix are obvious.  We must spend a few moments to consider the density matrices of paired UHF and paired GHF.  Note that paired UHF requires an equal number of $\uparrow$-spin and $\downarrow$-spin electrons, while paired GHF requires an even number of electrons.

The paired UHF molecular orbital coefficients satisfy
\begin{equation}
\mathbf{C}_\mathrm{occ} = \begin{pmatrix} \mathbf{A} & \mathbf{0} \\ \mathbf{0} & \mathbf{A}^\star \end{pmatrix}
\end{equation}
so the density matrix is
\begin{align}
\bm{\gamma}
 &= \mathbf{C}_\mathrm{occ} \, \mathbf{C}^\dagger_\mathrm{occ}
\\
 &= \begin{pmatrix} \mathbf{A} \, \mathbf{A}^\dagger & \mathbf{0} \\ \mathbf{0} & \mathbf{A}^\star \, \mathbf{A}^\mathsf{T} \end{pmatrix}.
\nonumber
\end{align}
Then
\begin{subequations}
\begin{align}
\mathbf{P}
 &= \frac{1}{2} \, \left(\mathbf{A} \, \mathbf{A}^\dagger + \mathbf{A}^\star \, \mathbf{A}^\mathsf{T}\right),
\\
\mathbf{M}^z
 &= \frac{1}{2} \, \left(\mathbf{A} \, \mathbf{A}^\dagger - \mathbf{A}^\star \, \mathbf{A}^\mathsf{T}\right).
\end{align}
\end{subequations}
Clearly, $\mathbf{P}$ is real and $\mathbf{M}^z$ is purely imaginary.

Similarly, for paired GHF the orbital coefficients are
\begin{equation}
\mathbf{C}_\mathrm{occ} = \begin{pmatrix} \mathbf{A} & \mathbf{B} \\ -\mathbf{B}^\star & \mathbf{A}^\star \end{pmatrix}
\end{equation}
so that the density matrix is
\begin{equation}
\bm{\gamma} = \begin{pmatrix} \mathbf{A} \, \mathbf{A}^\dagger + \mathbf{B} \, \mathbf{B}^\dagger & -\mathbf{A} \, \mathbf{B}^\mathsf{T} + \mathbf{B} \, \mathbf{A}^\mathsf{T} \\ -\mathbf{B}^\star \, \mathbf{A}^\dagger + \mathbf{A}^\star \, \mathbf{B}^\dagger & \mathbf{A}^\star \, \mathbf{A}^\mathsf{T} + \mathbf{B} \, \mathbf{B}^\mathsf{T} \end{pmatrix}.
\end{equation}
Then the charge and spin density matrices are
\begin{subequations}
\begin{align}
\mathbf{P}
 &= \frac{1}{2} \, \left(\mathbf{A} \, \mathbf{A}^\dagger + \mathbf{B} \, \mathbf{B}^\dagger + \mathbf{A}^\star \, \mathbf{A}^\mathsf{T} + \mathbf{B} \, \mathbf{B}^\mathsf{T}\right),
\\
\mathbf{M}^x
 &= \frac{1}{2} \, \left(-\mathbf{B}^\star \, \mathbf{A}^\dagger + \mathbf{A}^\star \, \mathbf{B}^\dagger  -\mathbf{A} \, \mathbf{B}^\mathsf{T} + \mathbf{B} \, \mathbf{A}^\mathsf{T}\right),
\\
\mathbf{M}^x
 &= \frac{1}{2 \, \mathrm{i}} \, \left(-\mathbf{B}^\star \, \mathbf{A}^\dagger + \mathbf{A}^\star \, \mathbf{B}^\dagger  + \mathbf{A} \, \mathbf{B}^\mathsf{T} - \mathbf{B} \, \mathbf{A}^\mathsf{T}\right),
\\
\mathbf{M}^z
 &= \frac{1}{2} \, \left(\mathbf{A} \, \mathbf{A}^\dagger + \mathbf{B} \, \mathbf{B}^\dagger - \mathbf{A}^\star \, \mathbf{A}^\mathsf{T} - \mathbf{B} \, \mathbf{B}^\mathsf{T}\right).
\end{align}
\end{subequations}
Again, it is clear that $\mathbf{P}$ is real and $\vec{\mathbf{M}}$ is purely imaginary.

Recall from our earlier discussions that if a density matrix component is purely imaginary, the corresponding magnetization density vector field component vanishes.  We thus see that paired UHF and paired GHF both have $\vec{m}(\vec{r}) = \vec{0}$.  This is physically sensible, in that paired UHF and paired GHF remain time-reversal invariant.  Only mean-field wave functions which break time-reversal symmetry can have non-zero magnetization density vector fields,\footnote{Note that even a restricted open-shell determinant, which remains a spin eigenfunctions, breaks time-reversal invariance.} just as only those which break complex conjugation symmetry can have non-coplanar magnetization density vector fields.

\begin{table}[t]
\caption{Constraints on densities and current densities for various kinds of single determinants.  If the entry is -- then the corresponding vector must vanish; if the entry is $\checkmark$ then the corresponding vector is not constrained.
\label{Tab:Classification2}}
\begin{ruledtabular}
\begin{tabular}{llllll}
Type of Determinant & $\vec{j}(\vec{r})$  &  $\vec{m}(\vec{r})$        &  $\vec{J}^x(\vec{r})$  &  $\vec{J}^y(\vec{r})$  &  $\vec{J}^z(\vec{r})$  \\
\hline
Real RHF            &  --                 &  --                        &  --                    &  --                    &  --                   \\
Complex RHF         &  $\checkmark$       &  --                        &  --                    &  --                    &  --                   \\
Real UHF            &  --                 &  $m(\vec{r}) \, \hat{z}$   &  --                    &  --                    &  --                   \\
Paired UHF          &  $\checkmark$       &  --                        &  --                    &  --                    &  $\checkmark$         \\
Complex UHF         &  $\checkmark$       &  $m(\vec{r}) \, \hat{z}$   &  --                    &  --                    &  $\checkmark$         \\
Real GHF            &  --                 &  $m_y(\vec{r}) = 0$        &  --                    &  $\checkmark$          &  --                   \\
Paired GHF          &  $\checkmark$       &  --                        &  $\checkmark$          &  $\checkmark$          &  $\checkmark$         \\
Complex GHF         &  $\checkmark$       &  $\checkmark$              &  $\checkmark$          &  $\checkmark$          &  $\checkmark$         \\
\end{tabular}
\end{ruledtabular}
\end{table}

Let us take one more brief digression.  In addition to the magnetization density vector field $\vec{m}(\vec{r})$ we can define three other relevant densities.  There is of course the familiar charge density
\begin{equation}
n(\vec{r}) = \sum_{\mu,\nu} \chi_\mu(\vec{r}) \, \chi_\nu(\vec{r}) \, P_{\mu\nu}.
\end{equation}
For purposes of classifying determinants, it is uninteresting.  We can also define the charge current density (see, e.g., Ref. \onlinecite{fukutome1981}),
\begin{equation}
\vec{j}(\vec{r}) = -\mathrm{i} \sum_{\mu,\nu} \left[\chi_\nu(\vec{r}) \, \nabla \chi_\mu(\vec{r}) - \chi_\mu(\vec{r}) \, \nabla \chi_\nu(\vec{r})\right] \, P_{\mu\nu}.
\end{equation}
Only the antisymmetric (and hence imaginary) component of $\mathbf{P}$ contributes to $\vec{j}$.  And we can define the spin current density
\begin{equation}
\vec{J}^k(\vec{r}) = -\mathrm{i} \sum_{\mu,\nu} \left[\chi_\nu(\vec{r}) \, \nabla \chi_\mu(\vec{r}) - \chi_\mu(\vec{r}) \, \nabla \chi_\nu(\vec{r})\right] \, M^k_{\mu\nu},
\end{equation}
where here $k$ indexes $x$, $y$, or $z$.  Again, only the imaginary components of $\mathbf{M}^k$ contribute to $\vec{J}^k$.  Table \ref{Tab:Classification2} relates the different types of determinants to different restrictions on the current density $\vec{j}$, magnetization density $\vec{m}$, and spin current density $\vec{J}^k$.

\begin{figure*}[t]
\includegraphics[width=0.9\textwidth]{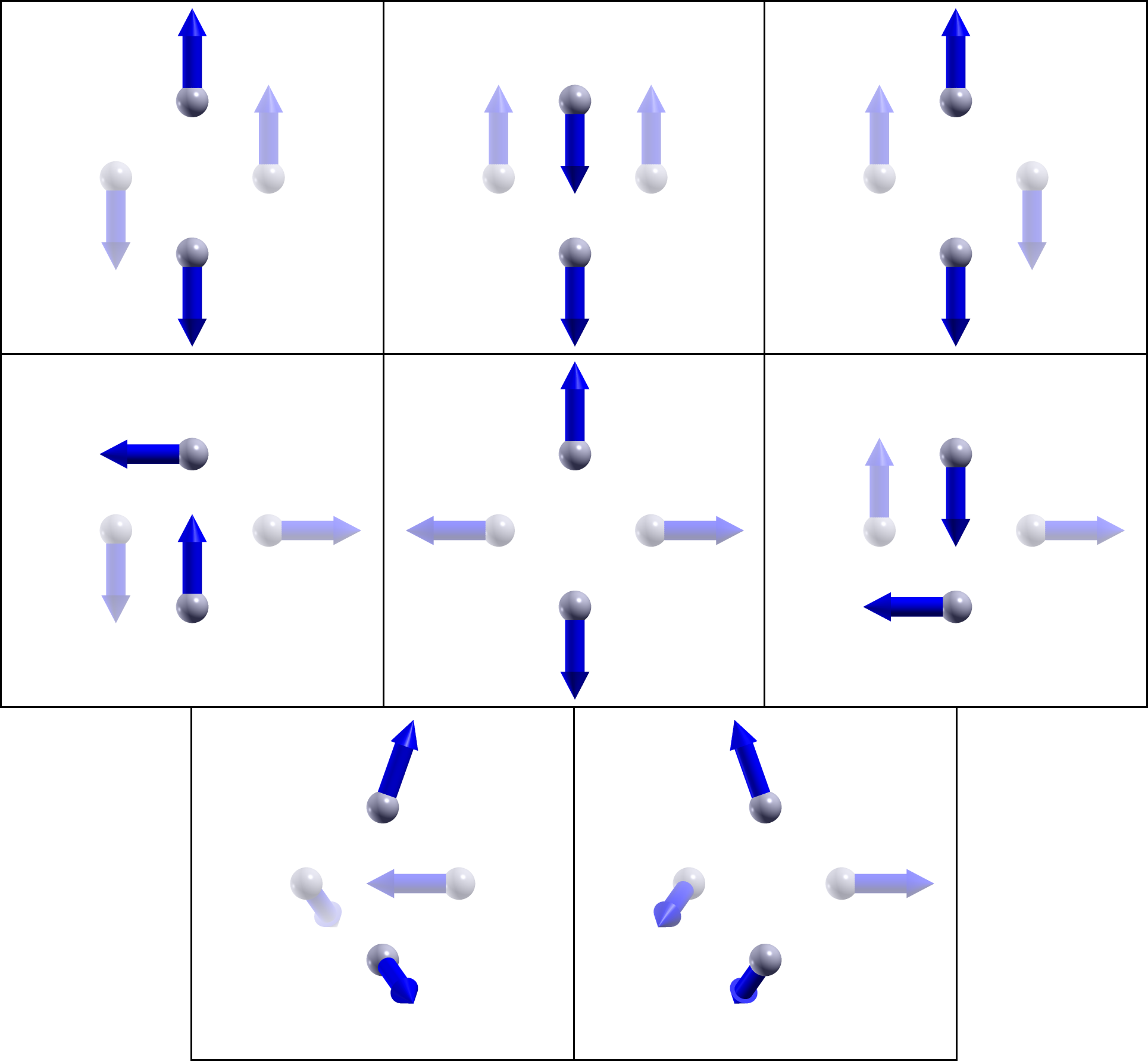}
\caption{Atomic magnetic moments in tetrahedral H$_4$.  Top row: the three distinct UHF solutions.  Middle row: the three distinct real GHF solutions.  Bottom row: the two distinct complex GHF solutions.
\label{Fig:H4Spins}}
\end{figure*}

\section{Applications}
Let us examine the basic idea with a few examples.

\subsection{Tetrahedral H$_4$}
Consider a uniform stretching of tetrahedral H$_4$.  We use the cc-pVDZ basis for simplicity.  Calculations are carried out in in-house code.  As the exact ground state is of singlet character for which $\langle \hat{\vec{S}} \rangle = \vec{0}$, we limit our discussion to Hartree-Fock solutions which satisfy this constraint.  We find a real RHF solution, a real UHF solution, a real GHF solution which following Ref. \onlinecite{Small2015} we will denote ``$\mathrm{rGHF}$'', and a complex GHF solution which we denote ``$\mathrm{cGHF}$.''  In fact, there are three distinct degenerate UHF solutions, three distinct degenerate real GHF solutions, and two distinct degenerate complex GHF solutions, the basic structures of which are shown in Fig. \ref{Fig:H4Spins}.  By ``distinct solutions'' we mean solutions which cannot be transformed into one another merely by spin rotation.

To initialize the GHF solutions, we add a Fermi contact term to the Hamiltonian and gradually turn off the strength of this perturbation.  Our Fermi contact perturbations are motivated by vibronic distortions of the electronic structure.  Tetrahedral H$_4$ is Jahn-Teller active; distorting the orbitals along the Jahn-Teller active modes\cite{Bersuker} without displacing the nuclei is thus likely to lead to lower energy solutions.  We should note that a global rotation of the Fermi contact bias is associated with a global rotation of the spin magnetization in the GHF solution.  Accordingly, the physically relevant quantity is the relative orientations of vectors on different atoms, but not the global orientation.  

\begin{figure}[t]
\includegraphics[width=0.95\columnwidth]{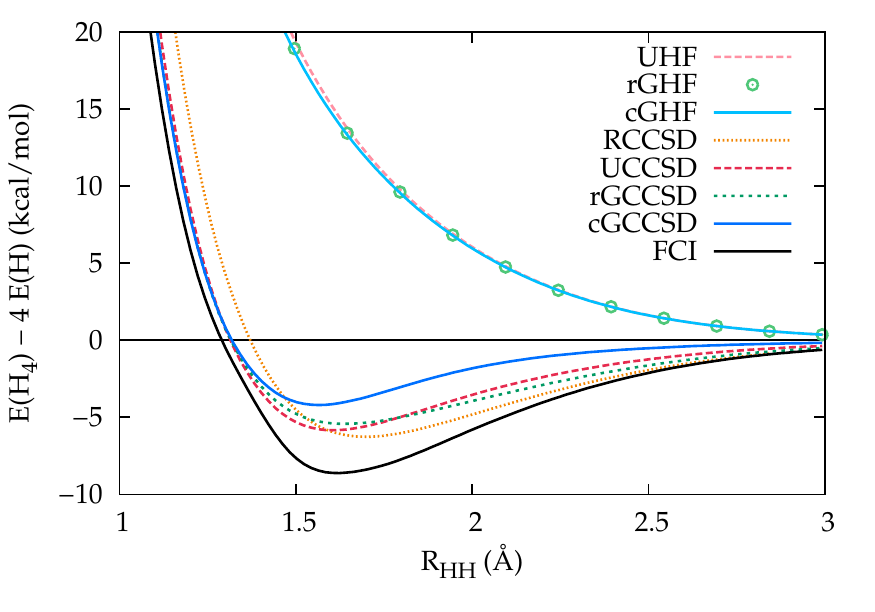}
\caption{Dissociation energies of tetrahedral H$_4$ in the cc-pVDZ basis set when uniformly stretched.  The RHF curve is far too high in energy to see on the plot.
\label{Fig:H4Energy}}
\end{figure}

In Fig. \ref{Fig:H4Energy} we show dissociation energies for the Hartree-Fock solutions as well as from coupled cluster with singles and doubles\cite{CCSD} (CCSD) based on these determinants.  We also show the full configuration interaction (FCI) curve as a reference.  Note that we have excluded RHF from the plot as RHF dissociates to the wrong limit and would not fit on our plot.  The other three Hartree-Fock solutions all dissociate correctly.  The real GHF is never more than a few milliHartree below the UHF, and the complex GHF is never more than a few milliHartree below the real GHF, so it is not easy to distinguish the various solutions on the plot.  Interestingly, the RHF-based CCSD is perhaps the best of the CCSD curves at large bond lengths, and the CCSD based on complex GHF is the worst of the lot.

We think it is valuable to understand the origin of the near-degeneracy between different spin arrangements for tetrahedral H$_4$. At long atomic separations, the Hamiltonian reduces to a Heisenberg Hamiltonian with an anti-ferromagnetic $J$. Interestingly, the frustration inherent in the tetrahedral arrangement yields an exact degeneracy in the HF solution to the Heisenberg Hamiltonian:\footnote{The exact ground state for tetrahedral H$_4$ and the equivalent Heisenberg Hamiltonian is also doubly degenerate.} the UHF solution with two spin-up and two spin-down electrons, the square planar arrangement, and the tetrahedral arrangement of spins all have the same energy. While in H$_4$ there are deviations from this degeneracy, they remain small and reflect discrepancies between the molecular Hamiltonian and the corresponding Heisenberg Hamiltonian which arise from our use of a finite interatomic separation.

\begin{figure}[t!]
\includegraphics[width=0.95\columnwidth]{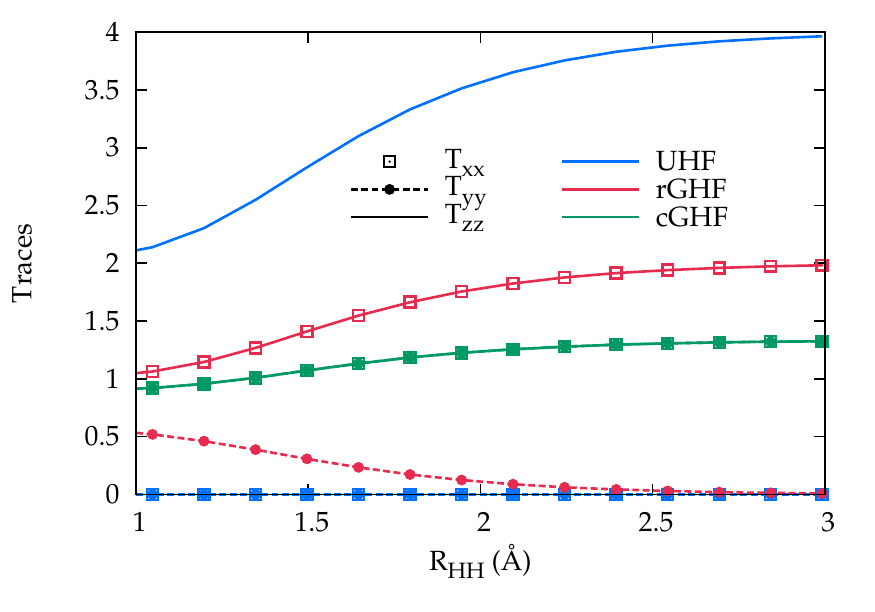}
\caption{Norms of the various magnetization density matrices for the three magnetic solutions in the symmetric dissociation of tetrahedral H$_4$.  Recall that, for example, $T_{xx} = \| \mathbf{M}^x \|^2 = \mathrm{Tr}(\mathbf{M}^x \, \mathbf{M}^x)$.  In this case, after rotation the real GHF has $\| \mathbf{M}^x \| = \| \mathbf{M}^z \|$ while for the complex GHF, all three components of $\vec{\mathbf{M}}$ have equal norms.
\label{Fig:H4Trace}}
\end{figure}

We are not, however, particularly interested in the total energies.  In Fig. \ref{Fig:H4Trace} we show the norms of the various components of $\vec{\mathbf{M}}$ after rotation for the three magnetic Hartree-Fock solutions.  Note that due to the high symmetry of the problem, all three components of $\vec{\mathbf{M}}$ have the same norm in the noncoplanar complex GHF, while $\mathbf{M}^x$ and $\mathbf{M}^z$ have the same norm for the coplanar real GHF.  We can tell that the complex GHF is noncoplanar by, for example, noticing that all three components of $\vec{\mathbf{M}}$ have non-zero real parts, or simply by checking $\bm{\mathcal{T}}$, which has three non-zero eigenvalues.  The noncoplanar GHF solution yields forces on the nuclei that respect the tetrahedral geometry, while the UHF and coplanar GHF solutions, in contract, are susceptible to a tetragonal Jahn-Teller distortion that can lower the energy.

\begin{figure*}[t]
\includegraphics[width=0.95\columnwidth]{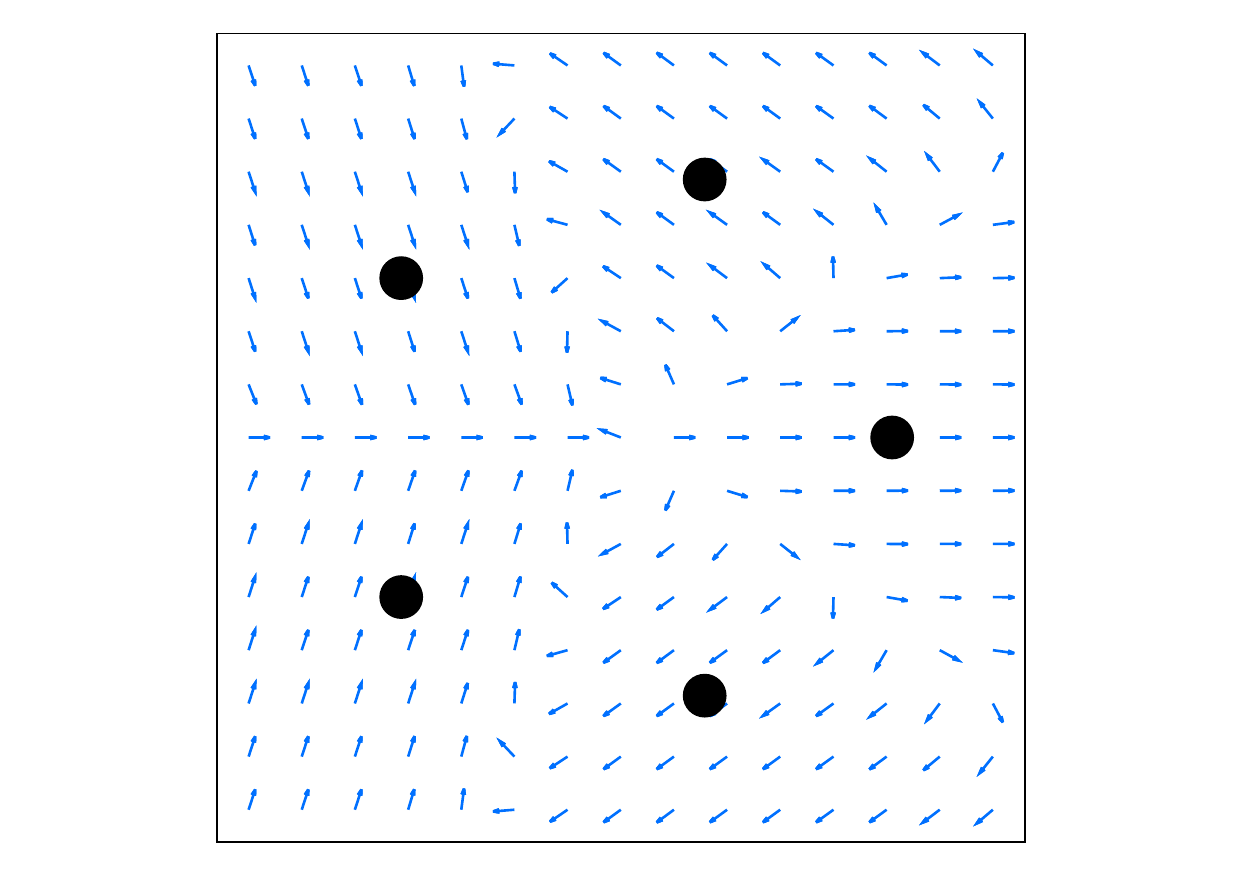}
\hfill
\includegraphics[width=0.95\columnwidth]{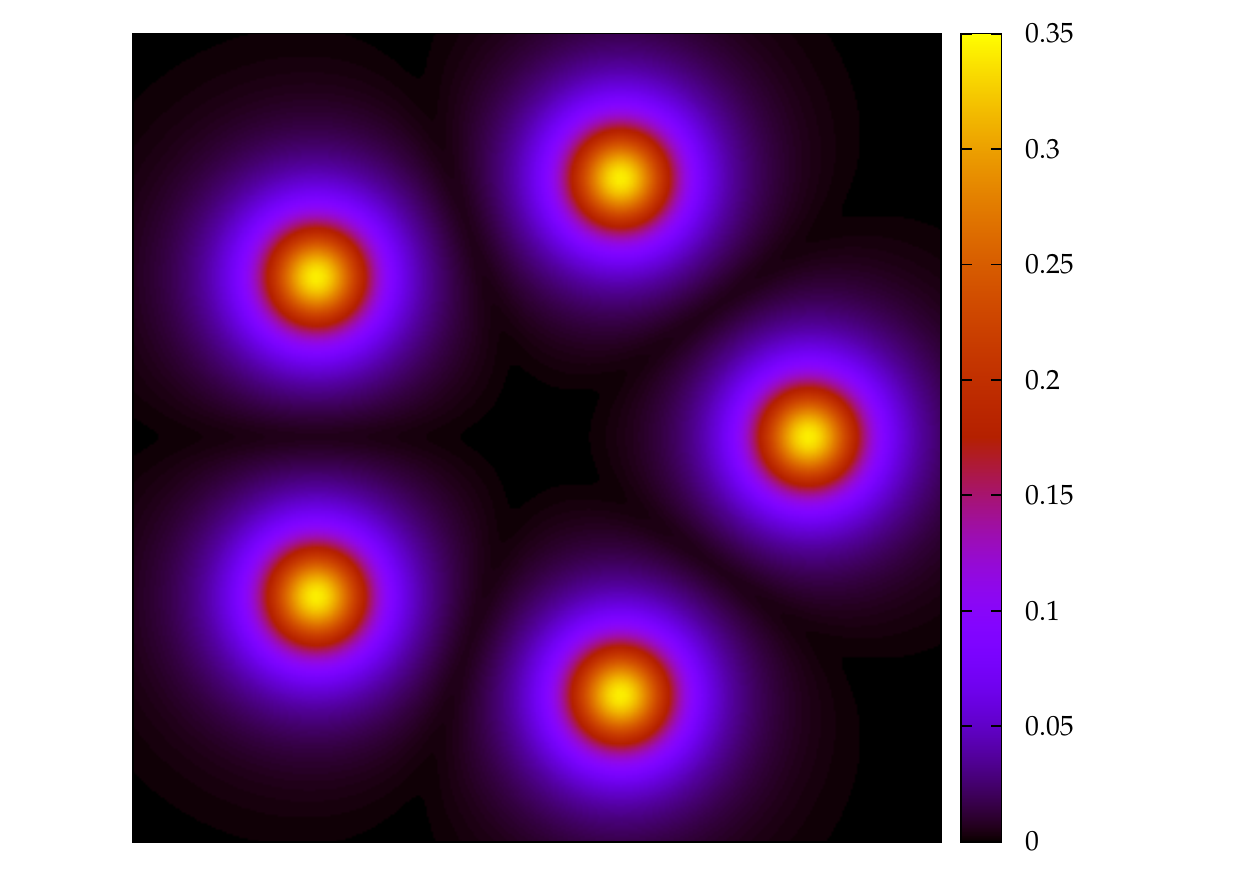}
\caption{Spin magnetization density vector field in the plane of the H$_5$ ring in the STO-3G basis set.  Left panel: unit vectors pointing in the direction of $\vec{m}(\vec{r})$.  Right panel: length of $\vec{m}(\vec{r})$.
\label{Fig:H5Plot}}
\end{figure*}

\subsection{Hydrogen Rings}
We next consider another artificial hydrogen system.  This time, we place five hydrogen atoms equally spaced around a circle such that the distance between nearest neighbor atoms is 3 Bohr.  For large interatomic separation, the hydrogen atoms should be coupled antiferromagnetically.  When the rings have an odd number of atoms, this leads to spin frustration and a GHF ground state\cite{Goings2015}.  We use the STO-3G basis set for maximal simplicity, and employ a Fermi contact term which directs the spin on each atom to be at an angle of $144^\circ$ from that on its neighbors.  We expect to converge to a coplanar GHF (and do; see Fig. \ref{Fig:H5Plot}).

To complicate things, and to showcase our coplanarity test, we do a global spin rotation of the Fermi contact term with arbitrary parameters $\alpha$, $\beta$, and $\gamma$ (see Eqn. \ref{Def:RotationOperators}).  The resulting wave function is complex (and in fact breaks complex conjugation symmetry).  After diagonalization, we find $T_{yy} = 0.156$ and $T_{xx} = T_{zz} = 1.713$, indicating a noncollinear solution.  Because the solution is noncollinear, we test for coplanarity, and after diagonalization we find $\mathcal{T}_{yy} = 0$ and $\mathcal{T}_{xx} = \mathcal{T}_{zz} = 1.713$, indicating coplanarity.  We do not generally expect the non-zero eigenvalues of $\mathbf{T}$ and of $\bm{\mathcal{T}}$ to be the same, but they are the same here because our determinant is just a spin rotation of a real GHF wave function for which $\mathbf{T}$ and $\bm{\mathcal{T}}$ have the same non-zero eigenvalues.  The spins in Fig. \ref{Fig:H5Plot} have been rotated back into the molecular plane.

\begin{figure*}[t]
\includegraphics[width=0.95\columnwidth]{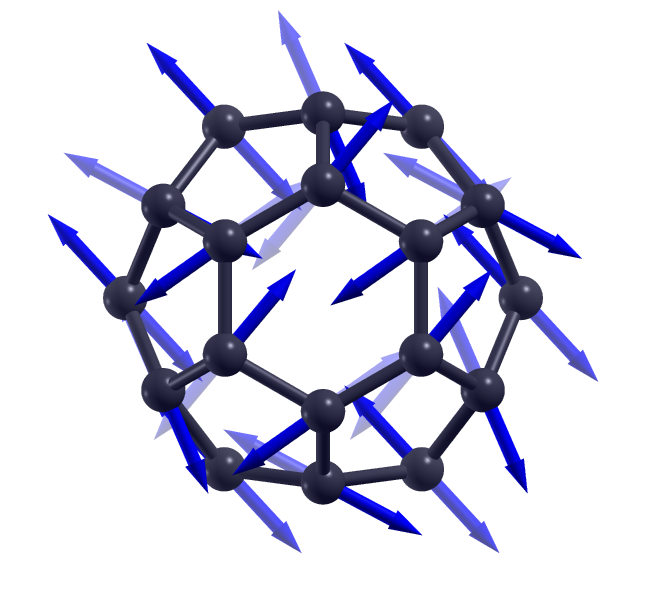}
\hfill
\includegraphics[width=0.95\columnwidth]{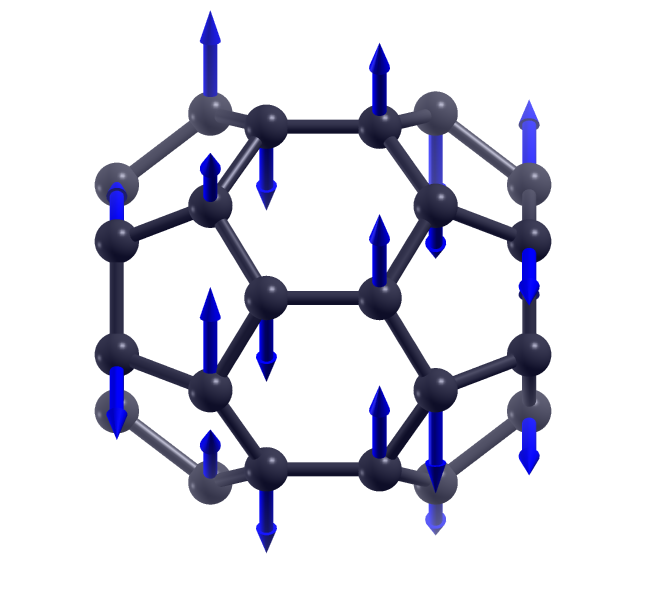}
\caption{Atomic magnetic moments in C$_{36}$.  Left panel: view from the cap.  Right panel: view from the hexacene ring belt.
\label{Fig:C36}}
\end{figure*}

\begin{figure*}[t]
\includegraphics[width=0.95\columnwidth]{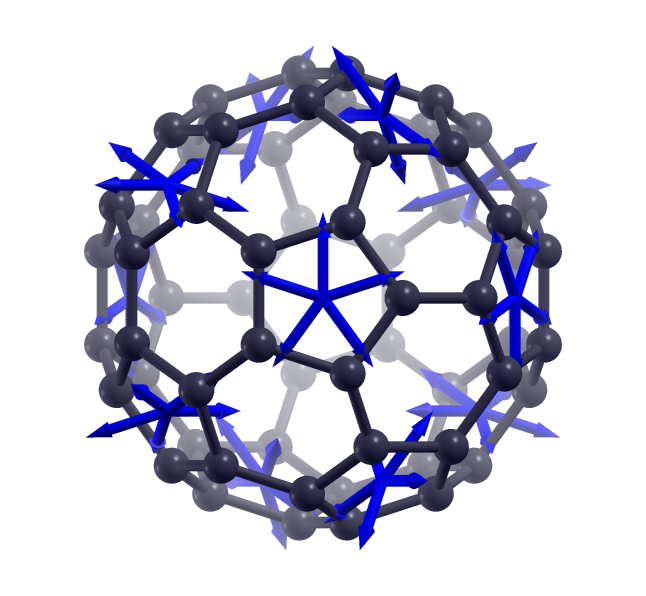}
\hfill
\includegraphics[width=0.95\columnwidth]{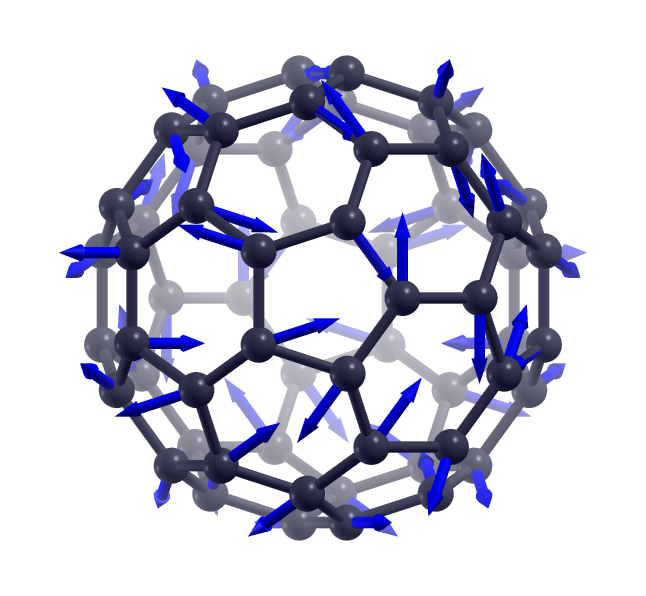}
\caption{Atomic magnetic moments in C$_{60}$.  The left panel shows the atomic magnetic moments translated to the centers of the pentagons, making the locally coplanar structure readily apparent.  The right panel shows the atomic magnetic moments themselves and emphasizes the antiferromagnetic character along hexagon edges.
\label{Fig:C60}}
\end{figure*}

\subsection{Fullerenes}
In previous work\cite{JimenezHoyos2014}, we have pointed out that there are non-collinear HF solutions for fullerene molecules. In simple terms, fused aromatic rings display a strong tendency towards anti-ferromagnetism \cite{Rivero2013}. In fullerenes, the presence of pentagon rings leads to frustration that is relieved by arranging the corresponding spins in non-collinear arrangements. Here, we choose to discuss two representative cases, namely C$_{36}$ and C$_{60}$.

The structure of C$_{36}$ (with $D_{6h}$ symmetry) can be thought of as an hexacene ring capped by an additional hexagon on the top and bottom. The GHF solution has all magnetic moments lying on the same plane, as illustrated in Fig. \ref{Fig:C36}.  There is full antiferromagnetic arrangement between carbon atoms in the hexacene ring related by a mirror plane perpendicular to the $C_6$ axis of the molecule.  In the case of C$_{60}$, the spin arrangement coincides with the one obtained by Coffey and Tugman\cite{Coffey1992} on the basis of the Heisenberg Hamiltonian. All atomic magnetic moments corresponding to the same pentagon are coplanar, but the planes corresponding to different pentagons are not parallel (left panel of Fig. \ref{Fig:C60}). There is exact antiferromagnetic arrangement along hexagon-hexagon edges (right panel of Fig. \ref{Fig:C60}).

While the magnetic structures discussed above were obtained by visual inspection, our collinearity tests fully confirms this picture, as seen in Tab. \ref{Tab:Fullerenes}.  In C$_{36}$ we have a coplanar solution (with $T_{zz} > T_{xx}$), while in C$_{60}$ we have a three-dimensional spin structure (with $T_{xx} = T_{yy} = T_{zz}$, leading to $\langle \hat{S}_x \rangle = \langle \hat{S}_y \rangle = \langle \hat{S}_z \rangle = 0$).

\begin{table}[t]
\caption{Eigenvalues of $\mathbf{T}$ and of $\bm{\mathcal{T}}$ for GHF solutions in C$_{36}$ and C$_{60}$.  From plotting magnetization densities we know that the magnetic structure in C$_{36}$ is noncollinear but coplanar, while in C$_{60}$ it is fully noncoplanar, as confirmed by our tests.
\label{Tab:Fullerenes}}
\begin{ruledtabular}
\begin{tabular}{ccc}
Eigenvalue         &  C$_{36}$   &   C$_{60}$   \\
\hline
$T_{xx}$            & 6.164 & 7.076 \\
$T_{yy}$            & 0.621 & 7.076 \\
$T_{zz}$            & 9.989 & 7.076 \\
\hline
$\mathcal{T}_{xx}$  & 6.164 & 6.761 \\
$\mathcal{T}_{yy}$  & 0.000 & 6.761 \\
$\mathcal{T}_{zz}$  & 9.989 & 6.761 \\
\end{tabular}
\end{ruledtabular}
\end{table}

\section{Conclusions}
Describing magnetic phenomena at a first-principles level is not always straightforward, even in the absence of an applied external magnetic field.  For correlated wave functions, magnetic ordering can be discerned by examining the two-particle density matrix or even higher-body density matrices.  The situation is simpler at the mean-field level, where the one-particle density matrix suffices.

Particularly at the mean-field level, the description of magnetism is frequently related to symmetry breaking.  Unfortunately, spin symmetry can break in manifold ways, and we would like a simple way to determine the form of symmetry breaking.  In general this requires considering the two-particle density matrix\cite{Small2015}, though again the one-particle density matrix is enough to understand the precise form of symmetry breaking for mean-field wave functions.  Even for correlated broken symmetry wave functions, there may be a significant amount of information to be gleaned from single-particle properties.

There are three main messages of this work.  The first is that noncoplanar magnetism requires an underlying complex conjugation symmetry breaking, just as non-zero spin magnetization density requires an underlying time-reversal symmetry breaking.  Second, a relatively straightforward examination of the one-particle density matrix can provide complete information about the magnetic structure of a single-determinant wave function and useful, albeit incomplete, information about the magnetic structure of a correlated wave function.  The noncollinearity test discussed here is equivalent to that of SSHG for single determinants; the coplanarity test is novel.  Finally, we want to reiterate that because the one-particle density matrix encapsulates all relevant information about single-determinant wave functions, one can readily see which symmetries a mean-field wave function has broken simply by looking at the density matrix without resort to orbital coefficients; indeed, the density matrix is perhaps a better place to look because unlike orbital coefficients, it is invariant to any orbital rotation which changes the wave function by no more than an overall phase.

\begin{acknowledgments}
This work was supported by the National Science Foundation, under award CHE-1462434. GES is a Welch Foundation Chair (C-0036).  CAJH acknowledges support by start-up funding from Wesleyan University.
\end{acknowledgments}

\appendix
\section{Complex Coplanar GHF}
We have said that coplanar spin densities do not necessarily correspond to real GHF determinants or even to those which can be rotated to be real.  Here we wish to provide a few simple examples showing that a coplanar spin density cannot necessarily be made to correspond to a real GHF determinant.

For a density matrix to correspond to a single determinant, it merely needs to be Hermitian and idempotent; as a consequence of the latter, it traces to the integer particle number\cite{Coleman1963}.  

Consider, then, the density matrix $\bm{\gamma}$ with components
\begin{subequations}
\begin{align}
\mathbf{P}
 &= \begin{pmatrix}  \frac{1}{2} & 0 & \mathrm{i} \, x \\ 0 & \frac{1}{2} & 0 \\ -\mathrm{i} \, x & 0 & \frac{1}{2} \end{pmatrix},
\\
\mathbf{M}^x
 &= \begin{pmatrix} -\frac{1}{4} & 0 & 0 \\ 0 & x & 0 \\ 0 & 0 & \frac{1}{4} \end{pmatrix},
\\
\mathbf{M}^y
 &= \mathbf{0},
\\
\mathbf{M}^z
 &= \begin{pmatrix} \frac{1}{4} & 0 & 0 \\ 0 & x & 0 \\ 0 & 0 & -\frac{1}{4} \end{pmatrix}.
\end{align}
\end{subequations}
This density matrix is Hermitian and for $x = \frac{1}{\sqrt{8}}$ it is also idempotent; it therefore corresponds to some single determinant.  Because $\mathbf{M}^y = \mathbf{0}$ it is clearly coplanar, and it is noncollinear since $\mathbf{M}^x$ and $\mathbf{M}^z$ are not multiples of one another.  Because $\mathbf{P}$ is complex and spin rotations do not change $\mathbf{P}$, it is clear that $\bm{\gamma}$ corresponds to an intrinsically complex coplanar GHF.

While it is clear that a complex charge density matrix guarantees a complex GHF, one can have an intrinsically complex coplanar GHF even when $\mathbf{P}$ is real.  Consider the density matrix $\bm{\gamma}$ with components
\begin{subequations}
\begin{align}
\mathbf{P} &= \begin{pmatrix} \frac{1}{2} + \lambda  &  0 \\ 0  & \frac{1}{2} - \lambda \end{pmatrix} = \frac{1}{2} \mathbf{1} + \lambda \, \bm{\sigma}^z,
\\
\mathbf{M}^x &= \begin{pmatrix} \lambda & 0 \\ 0 & \lambda \end{pmatrix} = \lambda \, \mathbf{1},
\\
\mathbf{M}^y &= \begin{pmatrix} 0 & -\mathrm{i} \lambda \\ \mathrm{i} \lambda & 0 \end{pmatrix} = \lambda \, \bm{\sigma}^y,
\\
\mathbf{M}^z &= \begin{pmatrix} 0 & \left(-1-\mathrm{i}\right) \lambda \\ \left(-1 + \mathrm{i}\right)\lambda & 0 \end{pmatrix} = \lambda \, \left(\bm{\sigma}^y - \bm{\sigma}^x\right).
\end{align}
\end{subequations}
Again, $\bm{\gamma}$ is Hermitian and for $\lambda^2 = \frac{1}{20}$ it is idempotent and thus corresponds to a single determinant.  Because $\mathbf{M}^y$ is purely imaginary, the magnetization density is coplanar.  One can readily verify that $\vec{\mathbf{M}}$ is noncollinear, most easily by noting that $\mathbf{M}^x$, $\mathbf{M}^y$, and $\mathbf{M}^z$ are not all multiples of the same matrix $\mathbf{Z}$.

Spin rotations can change $\bm{\gamma}$ to $\tilde{\bm{\gamma}}$ and components $\mathbf{M}^k$ to $\tilde{\mathbf{M}}^k$ via an orthogonal transformation.  If $\tilde{\bm{\gamma}}$ is to be purely real, then $\tilde{\mathbf{M}}^y$ must be purely imaginary.  Since $\mathbf{M}^x$ and $\mathbf{M}^z$ have real parts on different matrix elements, a purely imaginary $\tilde{\mathbf{M}}^y$ can have no contributions from $\mathbf{M}^x$ or $\mathbf{M}^z$.

Because orthogonal transformations preserve the angles between vectors, if neither $\mathbf{M}^x$ nor $\mathbf{M}^z$ contributes to $\tilde{\mathbf{M}}^y$, then $\mathbf{M}^y$ contributes to neither $\tilde{\mathbf{M}}^x$ nor $\tilde{\mathbf{M}}^z$.  For $\tilde{\bm{\gamma}}$ to be real, both these matrices must be real, which means neither can have any contribution from $\mathbf{M}^z$.  But if no part of $\tilde{\bm{\gamma}}$ has any contribution from $\mathbf{M}^z$, the transformation could not have been invertible, let alone orthogonal.

In short, then, there is no spin rotation which can make $\tilde{\bm{\gamma}}$ real, yet the magnetization vector field $\vec{m}(\vec{r})$ is clearly coplanar.  A coplanar $\vec{m}(\vec{r})$ can arise from a density matrix $\bm{\gamma}$ which cannot be transformed to a real GHF by spin rotations.

\bibliography{GHF}

\begin{thebibliography}{19}%
\makeatletter
\providecommand \@ifxundefined [1]{%
 \@ifx{#1\undefined}
}%
\providecommand \@ifnum [1]{%
 \ifnum #1\expandafter \@firstoftwo
 \else \expandafter \@secondoftwo
 \fi
}%
\providecommand \@ifx [1]{%
 \ifx #1\expandafter \@firstoftwo
 \else \expandafter \@secondoftwo
 \fi
}%
\providecommand \natexlab [1]{#1}%
\providecommand \enquote  [1]{``#1''}%
\providecommand \bibnamefont  [1]{#1}%
\providecommand \bibfnamefont [1]{#1}%
\providecommand \citenamefont [1]{#1}%
\providecommand \href@noop [0]{\@secondoftwo}%
\providecommand \href [0]{\begingroup \@sanitize@url \@href}%
\providecommand \@href[1]{\@@startlink{#1}\@@href}%
\providecommand \@@href[1]{\endgroup#1\@@endlink}%
\providecommand \@sanitize@url [0]{\catcode `\\12\catcode `\$12\catcode
  `\&12\catcode `\#12\catcode `\^12\catcode `\_12\catcode `\%12\relax}%
\providecommand \@@startlink[1]{}%
\providecommand \@@endlink[0]{}%
\providecommand \url  [0]{\begingroup\@sanitize@url \@url }%
\providecommand \@url [1]{\endgroup\@href {#1}{\urlprefix }}%
\providecommand \urlprefix  [0]{URL }%
\providecommand \Eprint [0]{\href }%
\providecommand \doibase [0]{http://dx.doi.org/}%
\providecommand \selectlanguage [0]{\@gobble}%
\providecommand \bibinfo  [0]{\@secondoftwo}%
\providecommand \bibfield  [0]{\@secondoftwo}%
\providecommand \translation [1]{[#1]}%
\providecommand \BibitemOpen [0]{}%
\providecommand \bibitemStop [0]{}%
\providecommand \bibitemNoStop [0]{.\EOS\space}%
\providecommand \EOS [0]{\spacefactor3000\relax}%
\providecommand \BibitemShut  [1]{\csname bibitem#1\endcsname}%
\let\auto@bib@innerbib\@empty
\bibitem [{\citenamefont {Fukutome}(1981)}]{fukutome1981}%
  \BibitemOpen
  \bibfield  {author} {\bibinfo {author} {\bibfnamefont {H.}~\bibnamefont
  {Fukutome}},\ }\href@noop {} {\bibfield  {journal} {\bibinfo  {journal} {Int.
  J. Quantum Chem.}\ }\textbf {\bibinfo {volume} {20}},\ \bibinfo {pages} {955}
  (\bibinfo {year} {1981})}\BibitemShut {NoStop}%
\bibitem [{\citenamefont {Stuber}\ and\ \citenamefont
  {Paldus}(2003)}]{stuber2003}%
  \BibitemOpen
  \bibfield  {author} {\bibinfo {author} {\bibfnamefont {J.~L.}\ \bibnamefont
  {Stuber}}\ and\ \bibinfo {author} {\bibfnamefont {J.}~\bibnamefont
  {Paldus}},\ }\enquote {\bibinfo {title} {Symmetry breaking in the independent
  particle model},}\ in\ \href@noop {} {\emph {\bibinfo {booktitle}
  {Fundamental World of Quantum Chemistry: A Tribute Volume to the Memory of
  Per-Olov L\"owdin}}},\ Vol.~\bibinfo {volume} {1},\ \bibinfo {editor} {edited
  by\ \bibinfo {editor} {\bibfnamefont {E.~J.}\ \bibnamefont {Br\"andas}}\ and\
  \bibinfo {editor} {\bibfnamefont {E.~S.}\ \bibnamefont {Kryachko}}}\
  (\bibinfo  {publisher} {Kluwer Academic Publishers},\ \bibinfo {address}
  {Dordrecht, The Netherlands},\ \bibinfo {year} {2003})\ Chap.~\bibinfo
  {chapter} {4}, pp.\ \bibinfo {pages} {67--139}\BibitemShut {NoStop}%
\bibitem [{\citenamefont {Jim\'enez-Hoyos}\ \emph {et~al.}(2011)\citenamefont
  {Jim\'enez-Hoyos}, \citenamefont {Henderson},\ and\ \citenamefont
  {Scuseria}}]{jimenezhoyos2011}%
  \BibitemOpen
  \bibfield  {author} {\bibinfo {author} {\bibfnamefont {C.~A.}\ \bibnamefont
  {Jim\'enez-Hoyos}}, \bibinfo {author} {\bibfnamefont {T.~M.}\ \bibnamefont
  {Henderson}}, \ and\ \bibinfo {author} {\bibfnamefont {G.~E.}\ \bibnamefont
  {Scuseria}},\ }\href@noop {} {\bibfield  {journal} {\bibinfo  {journal} {J.
  Chem. Theory Comput.}\ }\textbf {\bibinfo {volume} {7}},\ \bibinfo {pages}
  {2667} (\bibinfo {year} {2011})}\BibitemShut {NoStop}%
\bibitem [{\citenamefont {Yamaguchi}\ \emph {et~al.}(1999)\citenamefont
  {Yamaguchi}, \citenamefont {Yamanaka}, \citenamefont {Nishino}, \citenamefont
  {Takano}, \citenamefont {Kitagawa}, \citenamefont {Nago},\ and\ \citenamefont
  {Yoshioka}}]{Yamaguchi1999}%
  \BibitemOpen
  \bibfield  {author} {\bibinfo {author} {\bibfnamefont {K.}~\bibnamefont
  {Yamaguchi}}, \bibinfo {author} {\bibfnamefont {S.}~\bibnamefont {Yamanaka}},
  \bibinfo {author} {\bibfnamefont {M.}~\bibnamefont {Nishino}}, \bibinfo
  {author} {\bibfnamefont {Y.}~\bibnamefont {Takano}}, \bibinfo {author}
  {\bibfnamefont {Y.}~\bibnamefont {Kitagawa}}, \bibinfo {author}
  {\bibfnamefont {H.}~\bibnamefont {Nago}}, \ and\ \bibinfo {author}
  {\bibfnamefont {Y.}~\bibnamefont {Yoshioka}},\ }\href@noop {} {\bibfield
  {journal} {\bibinfo  {journal} {Theor. Chem. Acc.}\ }\textbf {\bibinfo
  {volume} {102}},\ \bibinfo {pages} {382} (\bibinfo {year}
  {1999})}\BibitemShut {NoStop}%
\bibitem [{\citenamefont {Small}\ \emph {et~al.}(2015)\citenamefont {Small},
  \citenamefont {Sundstrom},\ and\ \citenamefont {Head-{G}ordon}}]{Small2015}%
  \BibitemOpen
  \bibfield  {author} {\bibinfo {author} {\bibfnamefont {D.~W.}\ \bibnamefont
  {Small}}, \bibinfo {author} {\bibfnamefont {E.~J.}\ \bibnamefont
  {Sundstrom}}, \ and\ \bibinfo {author} {\bibfnamefont {M.}~\bibnamefont
  {Head-{G}ordon}},\ }\href {\doibase 10.1063/1.4913740} {\bibfield  {journal}
  {\bibinfo  {journal} {J. Chem. Phys.}\ }\textbf {\bibinfo {volume} {142}},\
  \bibinfo {pages} {094112} (\bibinfo {year} {2015})}\BibitemShut {NoStop}%
\bibitem [{\citenamefont {Jim\'enez-Hoyos}\ \emph {et~al.}(2012)\citenamefont
  {Jim\'enez-Hoyos}, \citenamefont {Henderson}, \citenamefont {Tsuchimochi},\
  and\ \citenamefont {Scuseria}}]{PHF}%
  \BibitemOpen
  \bibfield  {author} {\bibinfo {author} {\bibfnamefont {C.~A.}\ \bibnamefont
  {Jim\'enez-Hoyos}}, \bibinfo {author} {\bibfnamefont {T.~M.}\ \bibnamefont
  {Henderson}}, \bibinfo {author} {\bibfnamefont {T.}~\bibnamefont
  {Tsuchimochi}}, \ and\ \bibinfo {author} {\bibfnamefont {G.~E.}\ \bibnamefont
  {Scuseria}},\ }\href@noop {} {\bibfield  {journal} {\bibinfo  {journal} {J.
  Chem. Phys.}\ }\textbf {\bibinfo {volume} {136}},\ \bibinfo {pages} {164109}
  (\bibinfo {year} {2012})}\BibitemShut {NoStop}%
\bibitem [{\citenamefont {Thouless}(1960)}]{Thouless1960}%
  \BibitemOpen
  \bibfield  {author} {\bibinfo {author} {\bibfnamefont {D.~J.}\ \bibnamefont
  {Thouless}},\ }\href@noop {} {\bibfield  {journal} {\bibinfo  {journal}
  {Nucl. Phys.}\ }\textbf {\bibinfo {volume} {21}},\ \bibinfo {pages} {225}
  (\bibinfo {year} {1960})}\BibitemShut {NoStop}%
\bibitem [{\citenamefont {Ring}\ and\ \citenamefont {Schuck}(1980)}]{Ring80}%
  \BibitemOpen
  \bibfield  {author} {\bibinfo {author} {\bibfnamefont {P.}~\bibnamefont
  {Ring}}\ and\ \bibinfo {author} {\bibfnamefont {P.}~\bibnamefont {Schuck}},\
  }\href@noop {} {\emph {\bibinfo {title} {The Nuclear Many-Body Problem}}}\
  (\bibinfo  {publisher} {Springer-Verlag},\ \bibinfo {address} {New York,
  NY},\ \bibinfo {year} {1980})\BibitemShut {NoStop}%
\bibitem [{\citenamefont {Blaizot}\ and\ \citenamefont
  {Ripka}(1985)}]{Blaizot85}%
  \BibitemOpen
  \bibfield  {author} {\bibinfo {author} {\bibfnamefont {J.-P.}\ \bibnamefont
  {Blaizot}}\ and\ \bibinfo {author} {\bibfnamefont {G.}~\bibnamefont
  {Ripka}},\ }\href@noop {} {\emph {\bibinfo {title} {Quantum Theory of Finite
  Systems}}}\ (\bibinfo  {publisher} {The MIT Press},\ \bibinfo {address}
  {Cambridge, MA},\ \bibinfo {year} {1985})\BibitemShut {NoStop}%
\bibitem [{\citenamefont {Weiner}\ and\ \citenamefont
  {Trickey}(1998)}]{Weiner1998}%
  \BibitemOpen
  \bibfield  {author} {\bibinfo {author} {\bibfnamefont {B.}~\bibnamefont
  {Weiner}}\ and\ \bibinfo {author} {\bibfnamefont {S.~B.}\ \bibnamefont
  {Trickey}},\ }\href@noop {} {\bibfield  {journal} {\bibinfo  {journal}
  {Intern. J. Quantum Chem}\ }\textbf {\bibinfo {volume} {69}},\ \bibinfo
  {pages} {451} (\bibinfo {year} {1998})}\BibitemShut {NoStop}%
\bibitem [{Note1()}]{Note1}%
  \BibitemOpen
  \bibinfo {note} {Note that even a restricted open-shell determinant, which
  remains a spin eigenfunctions, breaks time-reversal invariance.}\BibitemShut
  {Stop}%
\bibitem [{\citenamefont {Bersuker}(2006)}]{Bersuker}%
  \BibitemOpen
  \bibfield  {author} {\bibinfo {author} {\bibfnamefont {I.~B.}\ \bibnamefont
  {Bersuker}},\ }\href@noop {} {\emph {\bibinfo {title} {The Jahn-Teller
  Effect}}}\ (\bibinfo  {publisher} {Cambridge University Press},\ \bibinfo
  {address} {Cambridge},\ \bibinfo {year} {2006})\BibitemShut {NoStop}%
\bibitem [{\citenamefont {Purvis}\ and\ \citenamefont {Bartlett}(1982)}]{CCSD}%
  \BibitemOpen
  \bibfield  {author} {\bibinfo {author} {\bibfnamefont {G.~D.}\ \bibnamefont
  {Purvis}}\ and\ \bibinfo {author} {\bibfnamefont {R.~J.}\ \bibnamefont
  {Bartlett}},\ }\href@noop {} {\bibfield  {journal} {\bibinfo  {journal} {J.
  Chem. Phys.}\ }\textbf {\bibinfo {volume} {76}},\ \bibinfo {pages} {1910}
  (\bibinfo {year} {1982})}\BibitemShut {NoStop}%
\bibitem [{Note2()}]{Note2}%
  \BibitemOpen
  \bibinfo {note} {The exact ground state for tetrahedral H$_4$ and the
  equivalent Heisenberg Hamiltonian is also doubly degenerate.}\BibitemShut
  {Stop}%
\bibitem [{\citenamefont {Goings}\ \emph {et~al.}(2015)\citenamefont {Goings},
  \citenamefont {Ding}, \citenamefont {Frisch},\ and\ \citenamefont
  {Li}}]{Goings2015}%
  \BibitemOpen
  \bibfield  {author} {\bibinfo {author} {\bibfnamefont {J.~J.}\ \bibnamefont
  {Goings}}, \bibinfo {author} {\bibfnamefont {F.}~\bibnamefont {Ding}},
  \bibinfo {author} {\bibfnamefont {M.~J.}\ \bibnamefont {Frisch}}, \ and\
  \bibinfo {author} {\bibfnamefont {X.}~\bibnamefont {Li}},\ }\href@noop {}
  {\bibfield  {journal} {\bibinfo  {journal} {J. Chem. Phys.}\ }\textbf
  {\bibinfo {volume} {142}},\ \bibinfo {pages} {154109} (\bibinfo {year}
  {2015})}\BibitemShut {NoStop}%
\bibitem [{\citenamefont {Jim\'enez-Hoyos}\ \emph {et~al.}(2014)\citenamefont
  {Jim\'enez-Hoyos}, \citenamefont {Rodr\'iguez-Guzm\'an},\ and\ \citenamefont
  {Scuseria}}]{JimenezHoyos2014}%
  \BibitemOpen
  \bibfield  {author} {\bibinfo {author} {\bibfnamefont {C.~A.}\ \bibnamefont
  {Jim\'enez-Hoyos}}, \bibinfo {author} {\bibfnamefont {R.~R.}\ \bibnamefont
  {Rodr\'iguez-Guzm\'an}}, \ and\ \bibinfo {author} {\bibfnamefont {G.~E.}\
  \bibnamefont {Scuseria}},\ }\href@noop {} {\bibfield  {journal} {\bibinfo
  {journal} {J. Phys. Chem. A}\ }\textbf {\bibinfo {volume} {118}},\ \bibinfo
  {pages} {9925} (\bibinfo {year} {2014})}\BibitemShut {NoStop}%
\bibitem [{\citenamefont {Rivero}\ \emph {et~al.}(2013)\citenamefont {Rivero},
  \citenamefont {Jim\'enez-Hoyos},\ and\ \citenamefont
  {Scuseria}}]{Rivero2013}%
  \BibitemOpen
  \bibfield  {author} {\bibinfo {author} {\bibfnamefont {P.}~\bibnamefont
  {Rivero}}, \bibinfo {author} {\bibfnamefont {C.~A.}\ \bibnamefont
  {Jim\'enez-Hoyos}}, \ and\ \bibinfo {author} {\bibfnamefont {G.~E.}\
  \bibnamefont {Scuseria}},\ }\href@noop {} {\bibfield  {journal} {\bibinfo
  {journal} {J. Phys. Chem. B}\ }\textbf {\bibinfo {volume} {117}},\ \bibinfo
  {pages} {12750} (\bibinfo {year} {2013})}\BibitemShut {NoStop}%
\bibitem [{\citenamefont {Coffey}\ and\ \citenamefont
  {Trugman}(1992)}]{Coffey1992}%
  \BibitemOpen
  \bibfield  {author} {\bibinfo {author} {\bibfnamefont {D.}~\bibnamefont
  {Coffey}}\ and\ \bibinfo {author} {\bibfnamefont {S.~A.}\ \bibnamefont
  {Trugman}},\ }\href@noop {} {\bibfield  {journal} {\bibinfo  {journal} {Phys.
  Rev. Lett.}\ }\textbf {\bibinfo {volume} {69}},\ \bibinfo {pages} {176}
  (\bibinfo {year} {1992})}\BibitemShut {NoStop}%
\bibitem [{\citenamefont {Coleman}(1963)}]{Coleman1963}%
  \BibitemOpen
  \bibfield  {author} {\bibinfo {author} {\bibfnamefont {J.}~\bibnamefont
  {Coleman}},\ }\href@noop {} {\bibfield  {journal} {\bibinfo  {journal} {Rev.
  Mod. Phys.}\ }\textbf {\bibinfo {volume} {35}},\ \bibinfo {pages} {668}
  (\bibinfo {year} {1963})}\BibitemShut {NoStop}%
\end{thebibliography}%

\end{document}